%% file: main.tex
\documentclass[journal]{IEEEtran}
\usepackage{placeins}

\usepackage{gensymb}

\usepackage{cite}
\usepackage[hidelinks,colorlinks=false]{hyperref}

\ifCLASSINFOpdf
   \usepackage[pdftex]{graphicx}
\else
   \usepackage[dvips]{graphicx}
\fi
\usepackage{epstopdf}
\graphicspath{{<pics>}}

\usepackage{algorithm}
\usepackage[noend]{algpseudocode}
\algdef{SE}[DOWHILE]{Do}{doWhile}{\algorithmicdo}[1]{\algorithmicwhile\ #1}%

\usepackage[colorinlistoftodos]{todonotes}
\usepackage{amsmath}
\usepackage{soul}

  \usepackage[caption=false,font=footnotesize]{subfig}
\usepackage{todonotes}
\newif\ifdraft
\drafttrue 
\usepackage{xcolor}

\newcommand{\heng}[1]{\ifdraft{\textcolor{black}{#1}}\fi}

\begin{document}
%
\title{ Hybrid Low-Power Wide-Area Mesh Network for IoT Applications}

%
%
%

\author{Xiaofan~Jiang,~\IEEEmembership{Member,~IEEE,}
        Heng~Zhang,~\IEEEmembership{Member,~IEEE,}
        Edgardo~Alberto~Barsallo~Yi,~\IEEEmembership{Member,~IEEE,}
        Nithin~Raghunathan,~\IEEEmembership{Member,~IEEE,}
        Charilaos Mousoulis,~\IEEEmembership{Member,~IEEE,}
        Somali~Chaterji,~\IEEEmembership{Member,~IEEE,}
        Dimitrios~Peroulis,~\IEEEmembership{Fellow,~IEEE,}
        Ali~Shakouri,~
        and~Saurabh~Bagchi,~\IEEEmembership{Fellow,~IEEE,}
\thanks{X. Jiang, H. Zhang, N. Raghunathan, D. Peroulis, A. Shakouri and S. Bagchi are with the School
of Electrical and Computer Engineering, Purdue University, West Lafayette,
IN, 47906 USA (e-mail: jiang175@purdue.edu.)}
\thanks{E. Yi is with the Department of Computer Science, Purdue University, West Lafayette, IN, 47906 USA.}
\thanks{S. Chaterji is with the School of Agricultural and Biological Engineering, Purdue University, West Lafayette, IN, 47906 USA.}
\thanks{Manuscript received April XX, XXXX; revised August XX, XXXX. (Xiaofan Jiang
and Heng Zhang are co-first authors.)}
}

%
%

\markboth{IEEE Internet of Things Journal, April 2020}%
{Shell \MakeLowercase{\textit{et al.}}: Bare Demo of IEEEtran.cls for IEEE Journals}
%



\maketitle
%
\IEEEpeerreviewmaketitle

\input{abstract}


\input{introduction}
\input{related}

\input{solution}

\input{results}

\input{limitation}

\input{conclusion}
\input{acknowledgement}

\FloatBarrier


\bibliographystyle{IEEEtran}
\bibliography{main}
%



%

\begin{IEEEbiography}[{\includegraphics[width=1in,height=1.25in,clip,keepaspectratio]{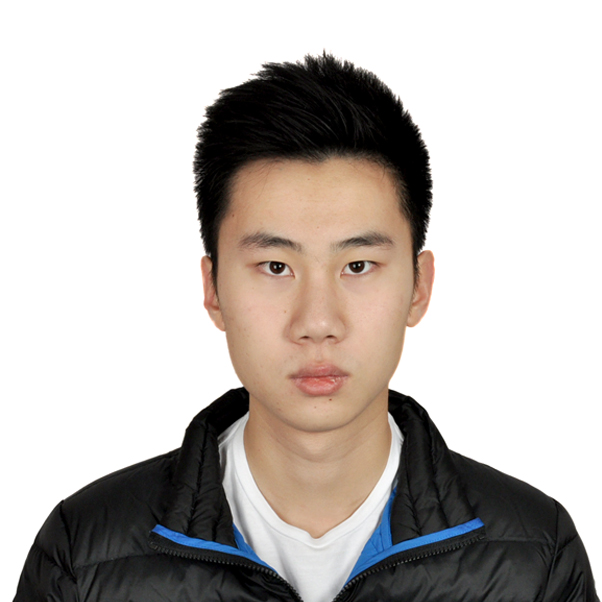}}]{Xiaofan Jiang}
is a Ph.D. Candidate at the School of Electrical and Computer Engineering, Purdue University in West Lafayette, Indiana. He is advised by Dimitrios Peroulis. He received the B.S. degree from Purdue University in 2015. His research interests include design and implementation of various embedded systems, wireless sensor networks and IoT devices for general and industry applications.
\end{IEEEbiography}

\begin{IEEEbiography}[{\includegraphics[width=1in,height=1.25in,clip,keepaspectratio]{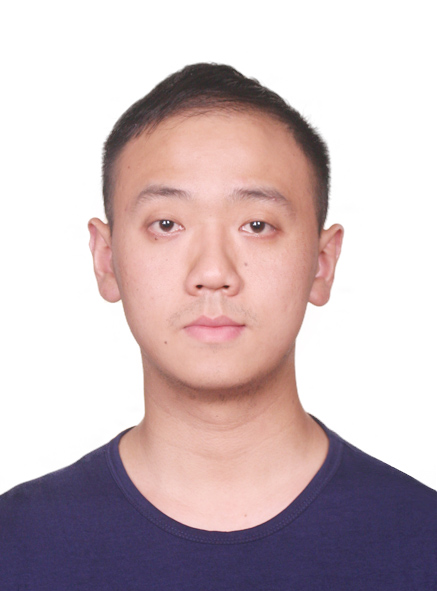}}]{Heng Zhang} is a Ph.D. Student at the School of Electrical and Computer Engineering, Purdue University West Lafayette, Indiana. He is advised by Saurabh Bagchi. He received the B.S. degree from Shanghai Jiao Tong University in 2016. His research interests include edge computing, mobile sensing, wearable resilience, and IoT networking. 
\end{IEEEbiography}

\begin{IEEEbiography}[{\includegraphics[width=1in,height=1.25in,clip,keepaspectratio]{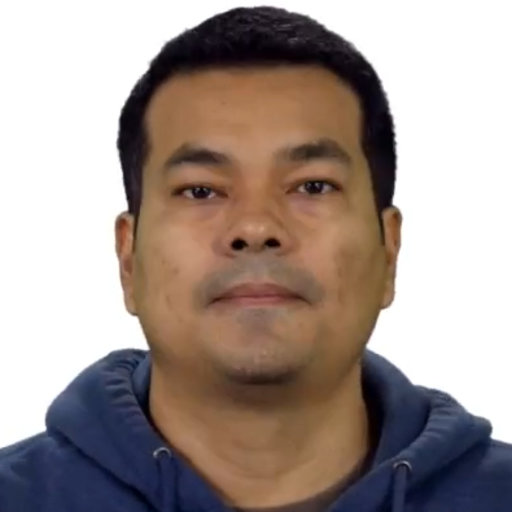}}]{Edgardo Barsallo Yi}
is a Ph.D. Candidate at the Computer Science Department, Purdue University. He is advised by Saurabh Bagchi. Before he joined Purdue, he worked as a software engineer and a software architect at the Indra Software Labs, Panama. He holds a Master in Software Engineering from UPSAM (Spain) and a B.S. in Computer Systems from UTP, Panama. His research interests include databases, distributed systems, mobile computing, and wearable resilience.
\end{IEEEbiography}

\begin{IEEEbiography}[{\includegraphics[width=1in,height=1.25in,clip,keepaspectratio]{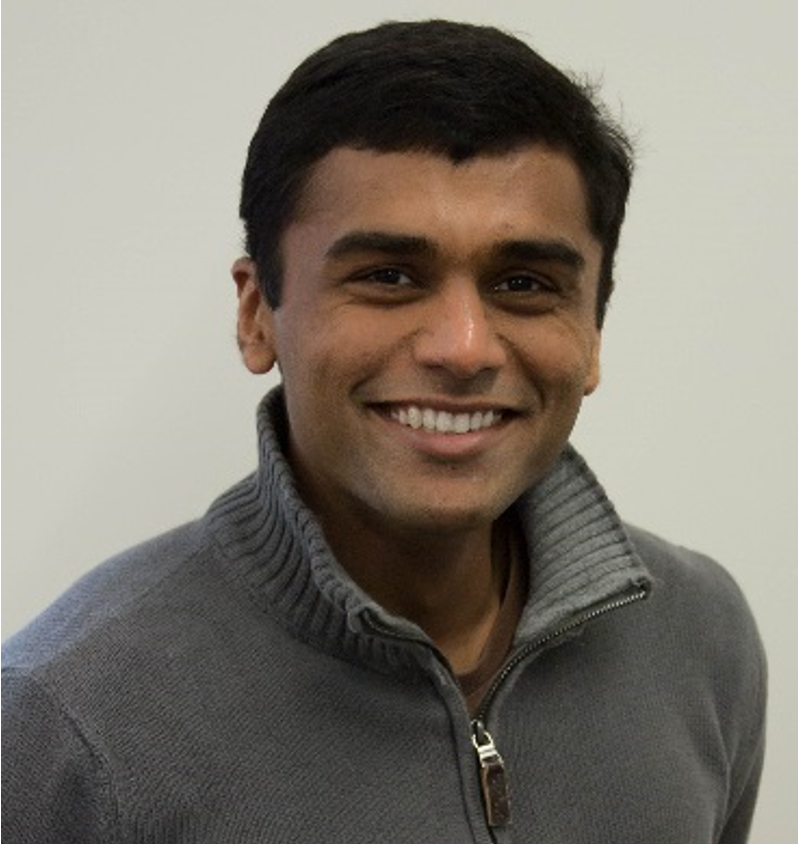}}]{Nithin Raghunathan}
received his Ph.D in electrical engineering from Purdue University, West Lafayette, IN, USA, in 2014. His dissertation focused on the development on micro-machined g-switches for impact applications typically in the ranges of 100 – 60,000 g’s. He worked as Post-Doctoral Research associate from 2014 to 2015 and was involved in the development of wireless radiation sensors for dosimetry applications. He is currently a Staff Scientist at the Birck Nanotechnology Center at Purdue University. He is currently working on Wabash Heartland innovation Network (WHIN) on the development of IoT sensors and network for Industrial and agricultural operations.  His other interests include novel MEMS inertial devices, development of new microfabrication techniques, wireless and flexible sensors and sensors for Lyophilization and aseptic processing and also sensors for industrial and harsh environments. 
\end{IEEEbiography}

\begin{IEEEbiography}[{\includegraphics[width=1in,height=1.25in,clip,keepaspectratio]{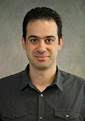}}]{Charilaos Mousoulis}
received the Ph.D. degree in Electrical and Computer Engineering from Purdue University, West Lafayette, IN, USA in 2012. Since 2014 he is a Senior Research Scientist at the School of Electrical and Computer Engineering, Purdue University, West Lafayette. He was previously a Postdoctoral Research Associate at the School of Biomedical Engineering, Purdue University, West Lafayette. His research interests include microsystems for biomedical applications, silicon-based radiation sensors for occupational dosimetry, sensors for food safety, flexible hybrid electronics, and IoT-based sensors for precision agriculture and advanced manufacturing.
\end{IEEEbiography}

\begin{IEEEbiography}[{\includegraphics[width=1in,height=1.25in,clip,keepaspectratio]{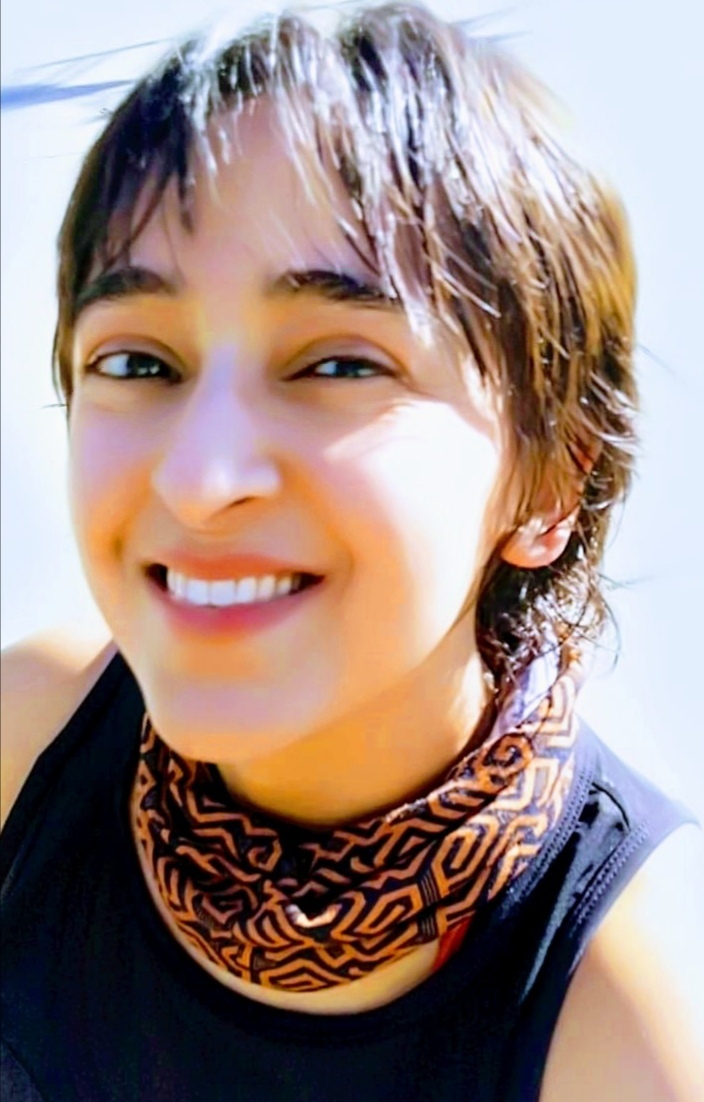}}]{Somali Chaterji}
 is an Assistant Professor in the Department of Agricultural and Biological Engineering at Purdue University, where she specializes in developing algorithms and statistical models for genome engineering, precision health, and digital agriculture. Dr. Chaterji got her PhD in Biomedical Engineering from Purdue University, winning the Chorafas International Award, College of Engineering Best Dissertation Award, and the Future Faculty Fellowship Award. She did her Post-doctoral Fellowship at the University of Texas at Austin in the Department of Biomedical Engineering, where her work was supported by an American Heart Association award. She followed this up with a Post-doctoral stint at Purdue Computer Science when she got her first NIH R01 on computational metagenomics. Dr. Chaterji is a technology commercialization enthusiast and has been consulting for the IC2 Institute at the University of Texas at Austin, since Spring 2014.
\end{IEEEbiography}

\begin{IEEEbiography}[{\includegraphics[width=1in,height=1.25in,clip,keepaspectratio]{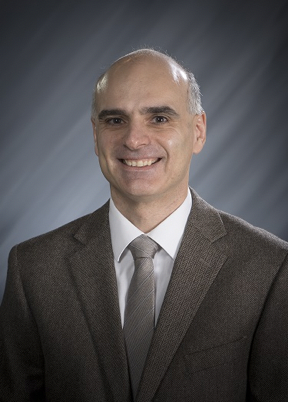}}]{Dimitrios Peroulis} (S’99–M’04–SM’15–F’17) is the Reilly Professor and Michael and Katherine Birck Head of the School of Electrical and Computer Engineering at Purdue University. He received his PhD degree in Electrical Engineering from the University of Michigan at Ann Arbor in 2003. His research interests are focused on the areas of reconfigurable systems, cold-plasma RF electronics, and wireless sensors. He has been a key contributor in developing high quality widely-tunable filters and novel filter architectures based on miniaturized high-Q cavity-based resonators in the 1-100 GHz range. He is currently leading research efforts in high-power multifunctional RF electronics based on cold-plasma technologies. He received the National Science Foundation CAREER award in 2008. He is an IEEE Fellow and has co-authored over 380 journal and conference papers. In 2019 he received the “Tatsuo Itoh” Award and in 2014 he received the Outstanding Young Engineer Award both from the IEEE Microwave Theory and Techniques Society (MTT-S). In 2012 he received the Outstanding Paper Award from the IEEE Ultrasonics, Ferroelectrics, and Frequency Control Society (Ferroelectrics section). His students have received numerous student paper awards and other student research-based scholarships. He has been a Purdue University Faculty Scholar and has also received ten teaching awards including the 2010 HKN C. Holmes MacDonald Outstanding Teaching Award and the 2010 Charles B. Murphy award, which is Purdue University's highest undergraduate teaching honor.
\end{IEEEbiography}

\begin{IEEEbiography}[{\includegraphics[width=1in,height=1.25in,clip,keepaspectratio]{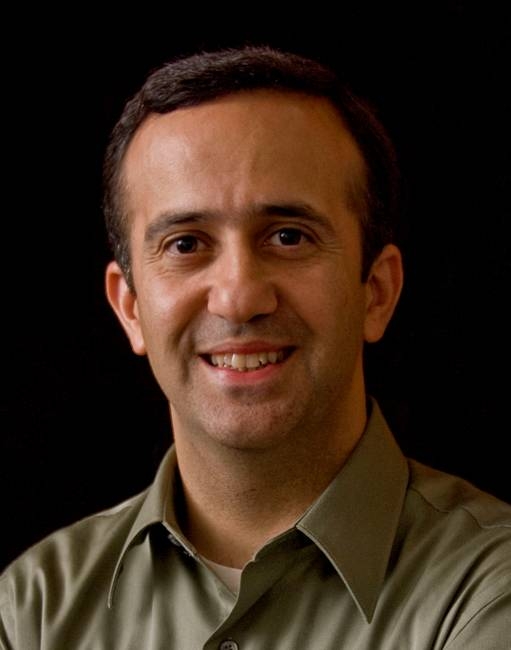}}]{Ali Shakouri}
 is the Mary Jo and Robert L. Kirk Director of the Birck Nanotechnology Center at Purdue University. He received his diplome d'Ingenieur in 1990 from Telecom ParisTech, France and his Ph.D. in 1995 from California Institute of Technology in Pasadena, CA. His group studies nanoscale heat transport and electrothermal energy conversion to improve electronic and optoelectronic devices. They have also developed novel imaging techniques to obtain thermal maps with sub diffraction-limit spatial resolution and 800ps time resolution. He is applying similar methods to enable real-time monitoring of functional film continuous manufacturing. He is leading SMART industry consortium to manufacture low-cost internet of thing (IoT) devices and sensor network. As a part of Wabash Heartland Innovation Network (WHIN), they are developing community IoT testbeds in advanced manufacturing and high tech agriculture.\end{IEEEbiography}

\begin{IEEEbiography}[{\includegraphics[width=1in,height=1.25in,clip,keepaspectratio]{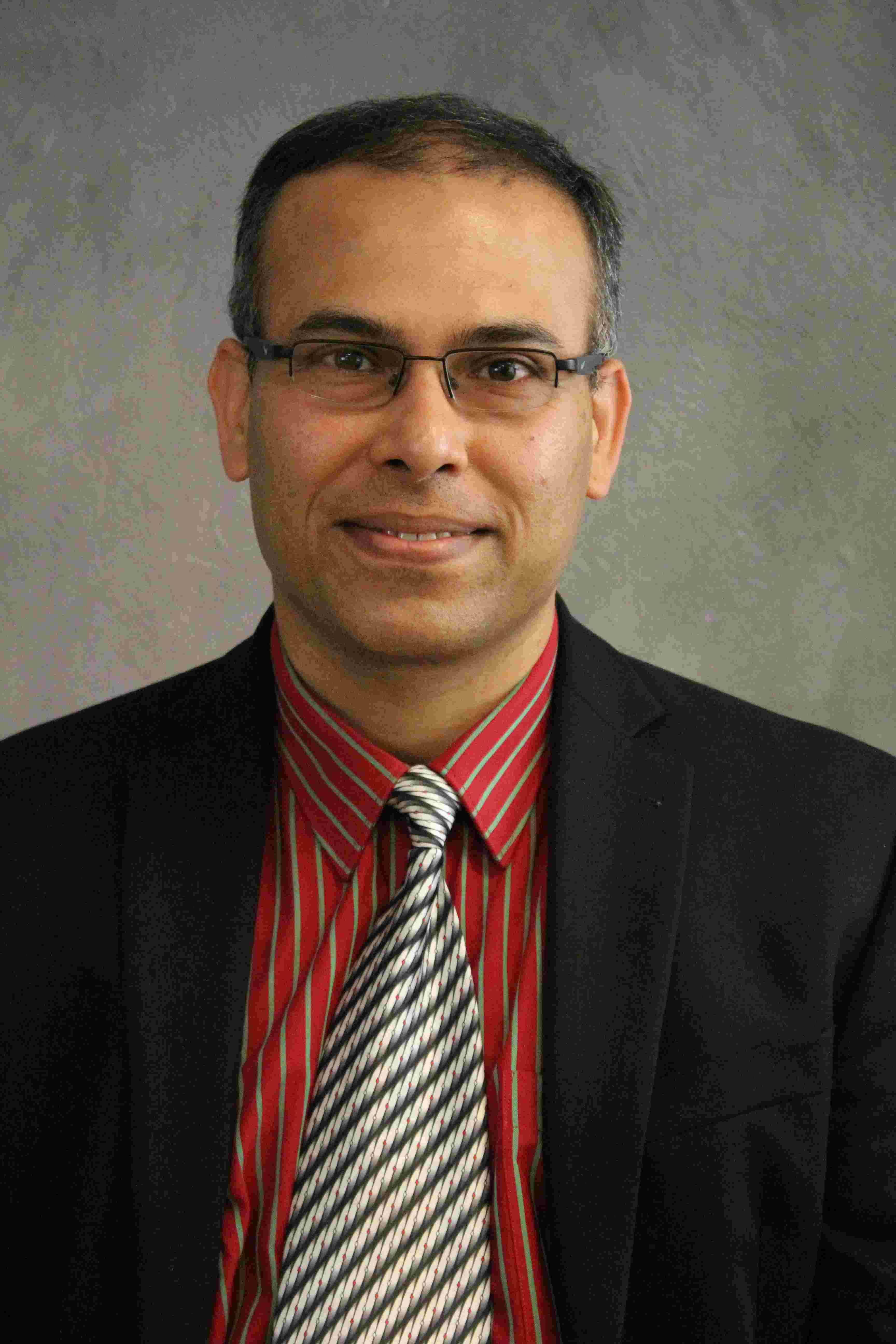}}]{Saurabh Bagchi}
is a Professor in the School of Electrical and Computer Engineering and the Department of Computer Science at Purdue University in West Lafayette, Indiana. He is the founding Director of a university-wide resilience center at Purdue called CRISP (2017-present). He was elected to the IEEE Computer Society Board of Governors for the 2017-19 term and re-elected in 2019. He is a co-lead on the WHIN-SMART center at Purdue for IoT and data analytics. Saurabh's research interest is in dependable computing and distributed systems. He is proudest of the 21 PhD students and 50 Masters thesis students who have graduated from his research group and who are in various stages of building wonderful careers in industry or academia. In his group, he and his students have way too much fun building and breaking real systems.
\end{IEEEbiography}
\vfill
\vfill
\vfill
\vfill






\end{document}

%% file: abstract.tex
\begin{abstract}
The recent advancement of the Internet of Things (IoT) enables the possibility of data collection from diverse environments using IoT devices. However, despite the rapid advancement of low-power communication technologies, the deployment of IoT networks still faces many challenges. In this paper, we propose a hybrid, low-power, wide-area network (LPWAN) structure that can achieve wide-area communication coverage and low power consumption on IoT devices by utilizing both sub-GHz long-range radio and 2.4 GHz short-range radio. Specifically, we constructed a low-power mesh network with LoRa, a physical-layer standard that can provide long-range (kilometers) point-to-point communication using custom time-division multiple access (TDMA). Furthermore, we extended the capabilities of the mesh network by enabling ANT, an ultra low-power, short-range communication protocol to satisfy data collection in dense device deployments. Third, we demonstrate the performance of the hybrid network with two real-world deployments at the Purdue University campus and at the university-owned farm. The results suggest that both networks have superior advantages in terms of cost, coverage, and power consumption vis-\`a-vis other IoT solutions, like LoRaWAN.

\end{abstract}

%% file: introduction.tex
\section{Introduction}
\label{sec:introduction}
%
%
%
%


As Internet of Things (IoT) devices, representing a network of interconnected things, proliferate with an estimated population of 125 billion IoT devices in the next decade, systems built out of these devices will play a role in many deployments, including those that need extended periods of unattended operation. 
These \textit{things} are essentially sensors and actuators, fitted with a wireless network interface, and computing and storage units.
IoT provides the connectivity to physically distributed devices, home appliances, and even devices in more critical sectors, such as healthcare, public utilities (e.g., electric grids), environmental monitoring, and transportation. These IoT devices sense, compute, and communicate, often in resource-limited deployments, forming a Wireless Sensor Network (WSN). In the smart city context, the IoT devices may monitor energy and utility distribution (e.g., smart grid); enable intelligent transportation systems, building automation, and smart homes. In the digital agriculture context, these devices may monitor various environmental conditions, such as soil moisture, nutrient quality, or microbial activity~\cite{digitalag2020}. The sensor data is collected and routed through the WSN and sent to the cloud for further analysis and possible closed-loop control. Our work addresses the development of a large-scale WSN that is suitable for distinct application areas, namely: \textit{digital agriculture} and \textit{smart and connected cities}. Our study is based on real-world deployments and highlights the practical design challenges and insights that arise from long-term unattended operation of those IoT systems. They highlight a distinct set of challenges in terms of the wireless networking capabilities. 





Digital agriculture refers to using modern technologies to increase the quantity and quality of agricultural products. Understanding the environmental conditions (e.g., temperature, humidity, and soil fertility) is important for agricultural management. Traditionally, it has been challenging to consistently monitor large farmlands due to the lack of automation and inefficient labor. Recent agricultural industries have been  adopting automation technologies that can monitor the environment and optimize farming~\cite{smart_agri_1, smart_agri_2, cps2019, digitalag2020}.
Some analytics can also be run on the back-end cloud computing resource and derive actionable knowledge from the raw data, such as, how to fertilize specific parts of the farm in a localized manner. Such kinds of decisions do not require real-time (or, near real-time) decision making. 

In addition to digital agriculture, smart city is another important application in IoT. It refers to using different types of sensors in urban areas to collect data and then use the data to manage urban assets and resources efficiently. According to the United Nations Department of Economic and Social Affairs, it is estimated that by 2050 there will be 2.5 billion people living in urban areas. WSNs are becoming part of a smart city infrastructure that can combat the strain of city growth by providing access to real-time data and can be realized through a robust network.

\begin{figure}[thbp]
\centering
\includegraphics[width=\linewidth]{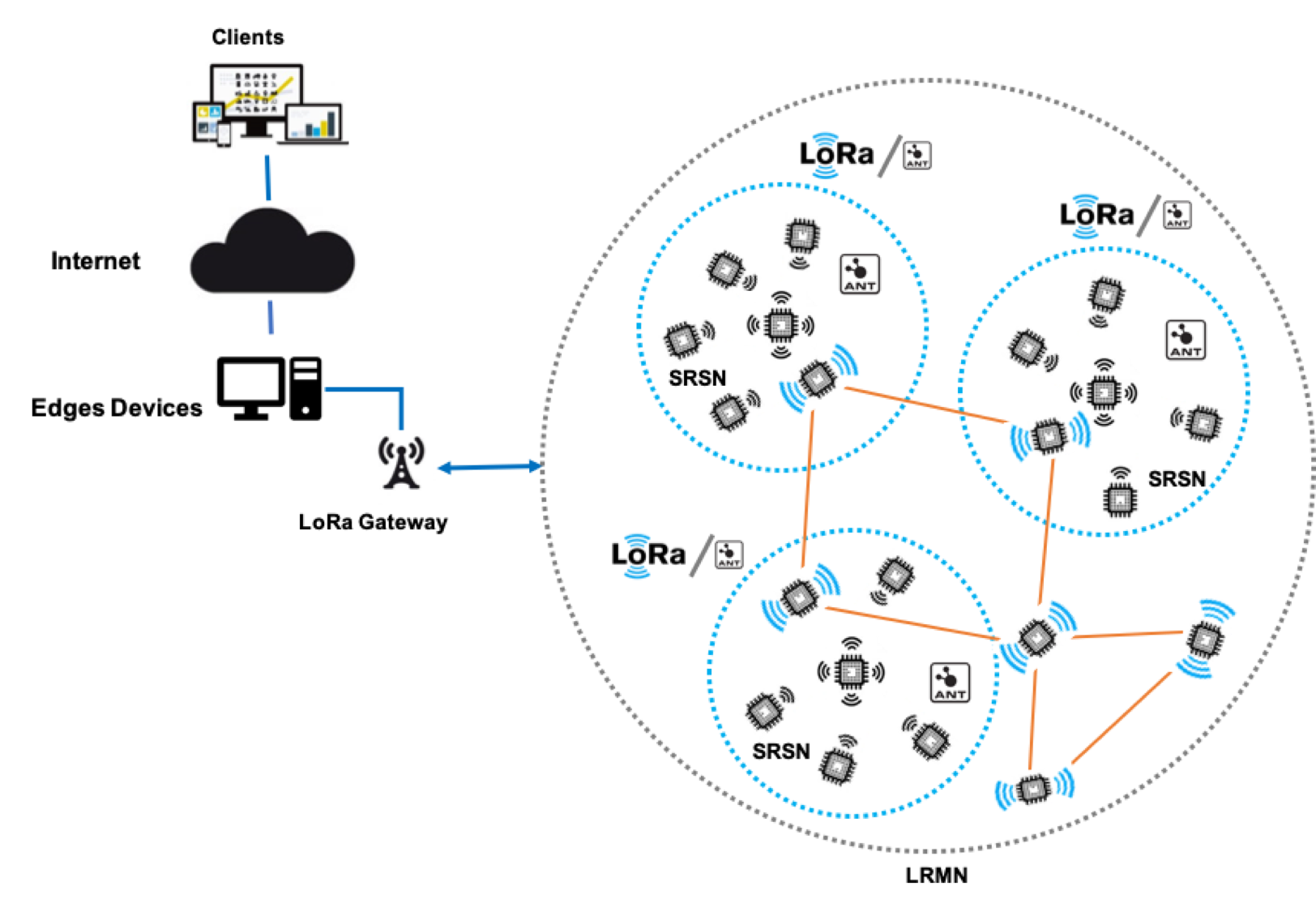}
\caption{Hybrid Network Architecture. Long Range Mesh Network (LRMN) consists of several Short Range Star Network (SRSN) and a few LoRa nodes. Each node is capable of communicating in a WPAN protocol (ANT in this case) or an LPWAN protocol (LoRa in this case). SRSN is used to satisfy dense sensing deployment while individual LoRa nodes are for sparse deployment.}
\label{fig:solution_architecture}
\end{figure}

\heng{However, there are several technical challenges to be solved for the successful deployment of such a WSN, such as harsh environmental conditions, communication range,  quality of service (QoS), deployment cost and energy consumption (battery life). These technical challenges are further discussed in our technical report~\cite{technical_report}.}
To handle the above challenges, we designed and fabricated a new custom wireless IoT hardware platform that is connected to temperature, humidity, and nitrate soil sensors and additional I/O pins for enabling the connections to other sensors. We choose LoRa~\cite{mikhaylov2016analysis} as the communication protocol because of its long range and low power consumption (Section~\ref{sec:LORA_ENERGY}). LoRa utilizes chirp spread spectrum (CSS) modulation and
operates in the sub-GHz ISM band to void penetration capability and heavy in-band interference. 
Furthermore, we proposed a lightweight, hybrid network combining the advantages of LoRa's wide area coverage and ANT's ultra-low power consumption by integrating them into a mesh network with following design goal:

\begin{itemize}
	\item Must be low-cost in system level, meaning not only hardware of the node is low-cost, the receiver (gateway) must be low-cost as well.
	\item Should be able to cover a large geographic area. For example, in our farming deployment, the network covers 2.2 $km^2$ of a farm and our campus deployment covers around 1.2 $km^2$ of Purdue campus.
	
	\item Has to be robust enough to survive the harsh environmental conditions and the network should able to handle node failures and recover from it.
	\item The deployment procedure for the IoT devices should not require any specific domain knowledge. Regular users should be able to simply install batteries to the IoT device and the device will work properly. 
	\item In addition to LoRa, the proposed network should also incorporate ultra-low power radio such as ANT to improve the performance and efficiency of the network in dense deployments.
	\item The proposed IoT device must be able to adopt to additional sensors.
\end{itemize}
    
    
To evaluate the performance of the proposed hybrid network, we conduct a series of experiments. First, we conduct several in-lab tests to show the power efficiency, deployment feasibility as well as the reliability of the network over time. Next, we deploy two real-world testbeds in both rural (Throckmorton-Purdue Agricultural Center--TPAC) and urban (West Lafayette campus at Purdue university) areas.

The rest of this paper is organized as follows. In Section~\ref{sec:related}, we provide a summary of popular WPANs and LPWANs technologies and discuss related works of LoRaWAN. In Section~\ref{sec:solution}, we present our solution with the network structure as well as the new hardware platform. The evaluation results are presented in Section~\ref{sec:results} and the paper is concluded with the discussions of limitations and future work in Section~\ref{sec:conclusion}.

%% file: related.tex
\section{Background and Related Works}
\label{sec:related}
\subsection{Summary of Current Technologies in WSN}

\begin{figure}[!ht]
\centering
\includegraphics[width=\linewidth]{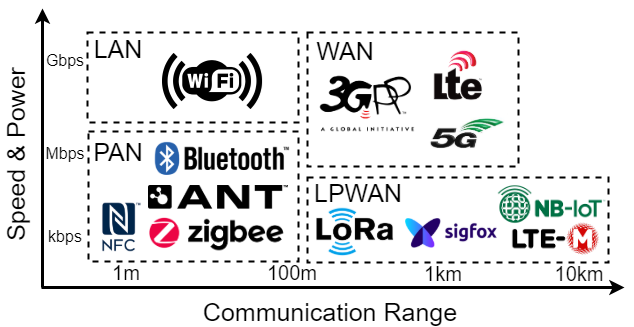}
\caption{A summary of current wireless commutation technologies, where trade-off between communication range (x-axis) and power/speed (y-axis) are clearly represented.}
\label{fig:wsns}
\end{figure}

Low power communication technologies for wireless IoT communication can grossly fall within two categories~(Fig. \ref{fig:wsns}): 


\begin{itemize}

  \item Wireless Personal Area Networks (WPANs): typically communicate from 10 meters to a few hundred meters. This category includes Bluetooth, Bluetooth Low Energy~(BLE), ANT, ZigBee, etc., which are applicable directly in short-range personal area networks or if designed in a mesh topology and with higher transmit power, larger area coverage is possible.
  
  \item Low-Power Wide Area Networks (LPWANs): have a communication range greater than one kilometer. Each gateway could communicate with thousands of end-devices. This category includes LoRaWAN, Sigfox, NB-IoT, etc. A summary of these technologies are summarised in Table~\ref{tab:related}~\cite{sigfox,LoRaWAN_spec,nokia}.
\end{itemize}

A more detailed discussion on the technologies can also be found in the technical report~\cite{technical_report}.
\begin{table*}[t]
\centering
\caption{Overview of LPWAN technologies: Sigfox, LoRa, and NB-IoT.}

\begin{tabular}{llll}
\hline
 & Sigfox\cite{sigfox} & LoRaWAN\cite{LoRaWAN_spec} & NB-IoT\cite{nokia} \\ \hline
Modulation & BPSK & CSS & QPSK \\
Band & \begin{tabular}[c]{@{}l@{}}Unlicensed ISM bands (868 MHz in\\ Europe, 915 MHz in North America)\end{tabular} & \begin{tabular}[c]{@{}l@{}}Unlicensed ISM bands (868 MHz in\\ Europe, 915 MHz in North America)\end{tabular} & \begin{tabular}[c]{@{}l@{}}Licensed LTE frequency\\ bands\end{tabular} \\
Bandwidth & 100Hz & 125 or 250kHz & 200kHz \\
Maximum data rate & 100bps & 50kbps & 200kbps \\
Bidirectional & Limited/Half-duplex & Yes/Half-duplex & Yes/Half-duplex \\
Maximum messages/day & 140(UL),4(DL) & Unlimited & Unlimited \\
Maximum payload length & 12 bytes (UL), 8 bytes (DL) & 243 bytes & 1600 bytes \\
Range & 10 km (urban), 40 km (rural) & 5 km (urban), 20 km (rural) & \begin{tabular}[c]{@{}l@{}}1 km (urban), 10 km\\ (rural)\end{tabular} \\
Interference immunity & Very high & Very high & Low \\
Authentication \& encryption & Not supported & Yes (AES 128b) & Yes (LTE encryption) \\
Adaptive data rate & No & Yes & No \\
Handover & End-devices do not join a single base station & End-devices do not join a single base station & \begin{tabular}[c]{@{}l@{}}End-devices join a\\ single base station\end{tabular} \\
Localization & Yes (RSSI) & Yes(TDoA) & No (under specification) \\
Allow private network & No & Yes & No \\
Standardization & No & LoRa-Alliance & 3GPP \\ \hline
\end{tabular}
\label{tab:related}
\end{table*}

In conclusion, we choose LoRa in our deployment because of the following advantages: i) the number of LoRa-enabled deployment is increasing continuously while, on the other hand, few initial NB-IoT deployments have been already deployed; ii) LoRa operates in the ISM band whereas cellular IoT operates in licensed bands; this fact favors the private LoRa networks without the involvement of mobile operators; iii) LoRaWAN, a cloud-based medium access control (MAC) layer protocol based on LoRa, has growing backing from industry, e.g.loRa Alliance, CISCO, IBM or HP, among others.



\subsection{Related Studies in LoRaWAN}
\label{sec:LORA_ENERGY}
The LoRaWAN network relies on the hub-and-spoke topology in which LoRaWAN gateways relay messages between end-devices and a central network server. This approach introduces two main problems: cost and power consumption. Deploying multiple gateways for a large LoRaWAN network is expensive, since LoRaWAN gateways normally cost from hundreds to thousands of dollars. In addition, LoRaWAN gateways require internet access to communicate with the server, which for many  applications, like smart agriculture, internet access might not be available.  In such cases, we will have to rely on cellular network which increases the network development cost.

The second issue is the power consumption. To achieve optimal transmission, LoRa utilises configuration parameters: the carrier frequency, the spreading factor, the bandwidth and the coding rate \cite{khutsoane2017iot}. The combination of these parameters affects energy consumption and transmission ranges. Taoufik et al.\cite{bouguera2018energy} calculate the theoretical maximum range that can be achieved at given output power ($P_{Tx}$) level with at different spreading factors (SF). In addition, they also proposed a energy consumption model based on these parameters as following:

\begin{equation}
\label{energy}
E_{tx} = \frac{P_{cons}(P_{Tx}) \cdot (N_{Payload} + N_{p} + 4.25) \cdot2^{SF}}{8\cdot PL\cdot BW}
\end{equation}

where $E_{tx}$ is the energy consumed per bit, $P_{cons}(P_{Tx})$ is the total consumed power which depends on transmission power ($P_{Tx}$), $PL$ is the payload size and BW is the bandwidth. 

\begin{figure}[!ht]
\centering
\includegraphics[width=\linewidth]{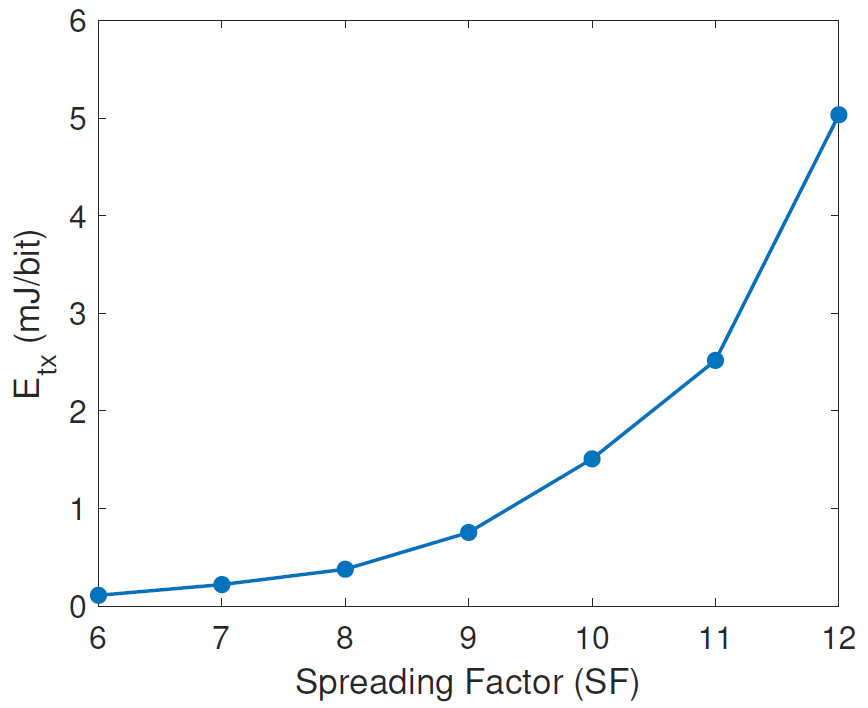}
\caption{LoRa Time-on-Air vs different SF with 8 bytes payload (CR = 4/5, BW = 125kHz and 8 preamble symbols)}
\label{fig:sf}
\end{figure}

To achieve long communication ranges with LoRaWAN ( $>$ 10 km), high spreading factor (SF) are required with a transmit power greater than 20 dBm (assuming a path-loss exponent equal to 3). However, a similar range can be achieved with 3 continuous hops using 3 different nodes with SF = 7. Lowering the spreading factor consumes significantly less energy. Total energy consumption for these two scenarios can be calculated based on Equation \ref{energy} for SX1262 LoRa transceivers at 20 dBm ($P_{cons}$(20dBm) = 389.4 mW \cite{sx1262}) with 8 bytes payload. Fig. \ref{fig:sf} shows the $E_{tx}$ for all spreading factors (6 to 12).



Furthermore, since we are talking about multi-hopping, the energy consumption for RX has to be considered as well and can be calculated as follows ($P_{rx}$ = 15.2 mW \cite{sx1262}):

\begin{equation}
SF= 7, E_{rx} = \frac{P_{rx} \cdot T_{on air}}{8} = \frac{15.2\cdot 36.096}{8*8} = 8.57 \text{uJ/bit}
\end{equation}


Hence, the total energy consumption for 3 hops is:

\begin{equation}
E_{Total} = 3 \cdot E_{tx} + 2 \cdot E_{rx} = 0.67 \text{mJ}
\end{equation}

Similar observations could be found for SF9, SF10, SF11 and SF12. As a conclusion, for a battery operated LoRa network covering a large range ($>$ 10$km$), a dynamic multi-hop mesh network could be much more efficient than LoRaWAN.
The power consumption is also distributed across multiple nodes, resulting in overall better life span for ``battery driven" WSN compared with LoRaWAN.

  In addition, LoRaWAN's asynchronous, ALOHA-based protocol limits its scalability and reliability\cite{zorbas2020ts}. Capacity of LoRaWAN networks are simulated and discussed in \cite{lorawan_limit}, which indicates LoRaWAN network has very limited capacity due to desiccation and duty-cycle restrictions. Varsier and Schwoerer \cite{varsier2017capacity} found that PDR reduced to 25\% due to packet collisions for a virtual large-scale application with high node densities. To overcome the limitations of LoRaWAN, more recent studies describe time-slot-based medium access mechanisms. While Piyare et al.\cite{piyare2018demand} describe an asynchronous time division multiple access (TDMA) with a separate wake-up radio channel (range of wake-up radio tested in lab environment, not multi-hop within sub-net), Reynders et al. \cite{reynders2018improving} suggest using lightweight scheduling that needs an adoption of the LoRaWAN network.

\subsection{Concurrent Transmission (CT) with LoRa}

Chun-Hao Liao \textit{et al}.~\cite{ctlora} developed a concurrent transmission~(CT) flooding based multi-hop LoRa network with low collision rate by introducing random delay. They demonstrated a successful deployment of 18 sensors between multiple buildings across 290 m x 195 m area. However, their approach falls short in the following two design rules of WSN.

\begin{figure}[!ht]
\centering
\includegraphics[width=\linewidth]{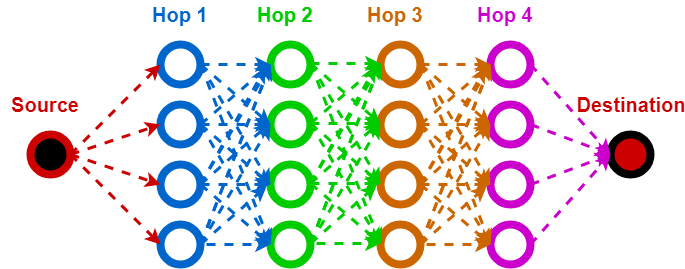}
\caption{Demonstration of LoRa Concurrent Transmission Problem \cite{ctlora}}
\label{fig:ct}
\end{figure}

First, the CT-flooding approach is not applicable for WSN due to high-power consumption. Figure~\ref{fig:ct} shows a basic relay map of a CT-LoRa network with 18 sensors. For the source node to transmit one data package to the destination node, this CT-LoRa will have a total of 17 transmission and 17 receiving windows across the network. While in a very optimized network only 5 transmission and receiving windows are required for transmitting the same package. This is especially true for any ``battery driven" LoRa based network since LoRa transmission is extremely expensive in terms of power when operating at high SF and transmission power.

Second, collision still exists with a high density of end devices. The authors studied the packet reception rate~(PRR) with different numbers of relays for each node. The results show that the PRR degrades significantly as the number of relays increases. These drawbacks limit the general scalability of their work in WSN.

\subsection{Synchronous LoRa mesh Network}

Recently, Ebi et al.\cite{ebi2019synchronous} proposed a mesh network approach to extend the capability of LoRaWAN by integration with a linear mesh network with multi-hopping to monitor underground infrastructure. The key feature of their network is the use of intermediate repeater nodes (RN) that allow the formation of individual linear multi-hop network with clusters of sensor nodes (SN). 
Although this network works for monitoring underground infrastructure, their implementation comes with some limitations. The bare-bone underlying structure is still LoRaWAN. As discussed in Section~\ref{sec:LORA_ENERGY}, 
the limitations of LoRaWAN still remain unsolved. In addition, the maximum number of the child-nodes is limited to 5 SNs due to the inherent payload restriction of the LoRaWAN standard. The author also observed that the RNs consume about twice as much energy as SNs because of the additional LoRaWAN communication with the gateway. This means the RNs will drain faster than SNs in battery life, resulting in a disproportionate failure rate. When RNs fail, all connected SNs will lose the network connection. Thus these limitations could result in higher failure rates in certain nodes and non-optimal power consumption.

%% file: solution.tex
\section{Our Solution}
\label{sec:solution}
To enable the data collection with varying sensors as well as to support wide area coverage with low energy consumption, we proposed a hybrid network with short and long-range communication links and designed our own sensor node by integrating low-power micro-controller with dual wireless communication interfaces (915MHz and 2.4GHz) to support the proposed network.

\subsection{Network Architecture}
Our hybrid network architecture is shown in Fig.~\ref{fig:solution_architecture}. Our network topology is a mesh of multiple smaller star-topology sub-networks. We use LoRa to build a \textbf{Long-Range Mesh Network} (LRMN) and ANT to build a \textbf{Short Range Star Network} (SRSN) for each individual sub-network. SRSN can cover a circular area of a radius of about 30 meters. SRSN works in the hub-and-spoke mode with a single hub node receiving data from multiple spoke nodes. 
There are two reasons behind the hybrid architecture. First, the IoT network should enable data collection in a wide area. Though LoRaWAN is capable of providing end-to-end communication of several miles, it suffers from high energy cost and the inadaptability of dynamic environments in applications like agriculture. Therefore, we utilize the long communication ability of LoRa to design the LRMN while we use the much more energy efficient SRSN for near-neighbor communication.
 Second, for certain applications, there may be some areas with dense deployments of sensor nodes, in which LoRa is an overkill and will cause network contention. Therefore, we design SRSN to collect data in such subareas for energy conservation.

\subsection{LoRa Mesh}

Our LoRa Mesh network supports the dynamic addition and removal of sensor node without causing other nodes to stop functioning or other manual efforts to reconfigure the network. During deployment, the new node only needs to be placed in the location of interest and it will join the mesh network automatically (Setup Phase in Algorithm~\ref{algo:lora_mesh}). 

\subsubsection{TDMA Scheduling Algorithm}
\label{sec:tdma}

One important requirement for the mesh network is to ensure the data is successfully uploaded to the cloud, no matter how far the sensor node is away from the LoRa gateway. Since the sensor node that is out of the communication range of a LoRa gateway needs to find a intermediate node for routing its data, the intermediate nodes and the sensor node need to coordinate the time window so that the intermediate node is in reception mode while the sensor node is sending its data. A trivial solution to this coordination problem is to always open the reception channel of the intermediate node. However, an always-on reception channel is energy inefficient (approximately 10 mA current for the device in LoRa reception mode vs 2 $\mu$A in sleep mode, 5000 times increase in power).

Therefore, we adopt a Time-Division Multiple Access (TDMA) scheduling algorithm~\cite{pantazis2009energy} and customize it to our system. \heng{The pseudo-code is shown in Algorithm~\ref{algo:tdma}. The input, $nodelist$ is the list of nodes in the network sorted in the descending hops to the LoRa gateway. Therefore, the algorithm starts with the node which is the furtherest from the gateway (line 7 in Algorithm~\ref{algo:tdma}) and ends with the node which is closest to the gateway.} There are three operating modes (Receive / Send / Sleep) of LoRa. The scheduling algorithm coordinates the sending and receiving actions for all nodes in the mesh network so that the data is transmitted without collisions with any other neighboring nodes that are also sending data at the same time. Additionally, customization puts the nodes mostly in Sleep mode. The difference between the original algorithm~\cite{pantazis2009energy} and \heng{Algorithm~\ref{algo:tdma}} is that the original algorithm assumes a node can send all the data, including its own sensor data as well as the data received from other nodes, in the same packet in one timeslot. \heng{However, each node can only send data with fixed size}. Therefore, if the data to be sent is too large, it has to be fragmented into multiple packets and sent at multiple timeslots. \heng{This customization is due to the packet size limitation in a single LoRa transmission. SX1262 LoRa transceiver~\cite{sx1262}, which we used in our deployment, has only 256 bytes of the transmission buffer}. Thus, assuming the size of a single fragmented data packet is 256 bytes, if a node has received 3 packets from 3 neighbors plus 1 packet of its own sensor data, it will need to send 4 packets with 1 KB size of data. The 1 KB data cannot be fulfilled in one timeslot due to the SX1262 buffer limitation. \heng{Line 22 of Algorithm~\ref{algo:tdma} checks the remaining packets (including those received from other nodes as well as the packets generated by itself) of a node and only after all of its packets have been scheduled, it will be removed from the nodelist so that the algorithm will not schedule it for the future timeslots. Section~\ref{sec:why_centralized} gives a concrete example to explain how the mesh network is built up using Algorithms~\ref{algo:tdma} and~\ref{algo:lora_mesh}}



\begin{algorithm}[h]
\caption{\heng{LoRa Communication Mode Scheduling}}\label{algo:tdma}
\begin{algorithmic}[1]
\Procedure{Build Schedule on the Hub Node}{}

\noindent\textbf{Input}: $nodelist$

\noindent\textbf{Output}: $schedule$
\Do
    \State $slot = new$ $Slot()$;
    \State $send\_coll\_list$ = [];
    \State $recv\_coll\_list$ = [];
    
    \While{$nodelist.length > 0$}
        \State $node=nodelist[0]$;
        \If{ ($node.recv==0$ 
             \State \textbf{and} 
             
             \State $node.packet>0$
             \State \textbf{and} 
             \State $!send\_coll\_list.contains(node)$ 
             \State \textbf{and} 
             \State $!receive\_coll\_list.contains(node)$) }
             
        \State $slot[node] = 'SEND'$;
        
        \State $slot[node.dest] = 'RECV'$;
        \State $send\_coll\_list.insert(node.dest.nbrs)$
        \State $send\_coll\_list.insert(node.dest)$
        \State $recv\_coll\_list.insert(node.nbrs)$
        \State $node.dest.recv --$;
        \State $node.packet --$;
        \If{ ($node.packet == 0$)}
        \State $nodelist.remove(node);$
        \EndIf
        \EndIf
    \EndWhile
    \State $schedule.insert(slot)$
  \doWhile{$slot.length >0$} 
  \State \textbf{return} $schedule$
\EndProcedure
\end{algorithmic}
\end{algorithm}

\subsubsection{Our Mesh Protocol}

\begin{figure}[!ht]
\centering
\includegraphics[width=\linewidth]{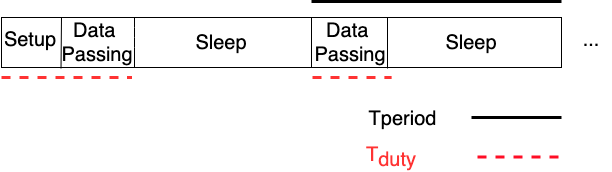}
\caption{Three phases of the nodes in LoRa mesh. Duty cycle consists of the initial Setup phase and any Data Passing phases. One period cycle consists of one Data Passing and the followed Sleep phase.}
\label{fig:two_phases}
\end{figure}

The LoRa node in our mesh network has three phases as shown in Figure~\ref{fig:two_phases}, namely \textit{Setup, Data Passing, } and \textit{Sleep}. The first phase (line 2 to 6 in Algorithm~\ref{algo:lora_mesh}) is the setup phase where the nodes build up a routing table and the LoRa hub node builds a communication schedule that indicates the time window for which nodes to send data \heng{(Algorithm~\ref{algo:tdma})}. The second phase (line 7 in Algorithm~\ref{algo:lora_mesh}) is the data passing phase where each node follows the schedule to send or receive data. In the third phase (line 8 in Algorithm~\ref{algo:lora_mesh}), the node sleeps to save energy since it knows no one will send data to it. 

\begin{algorithm}[h]
\caption{LoRa Mesh Network Protocol}\label{algo:lora_mesh}
\begin{algorithmic}[1]
\Procedure{Operations of Each Node}{}

 \textbf{Setup Phase} 
 
\State \heng{Each node builds local routing table using distance vector routing protocol}.
\State Each node broadcasts its own routing table and forwards others' routing tables, \heng{for the purpose of gathering all routing tables at the Hub node.}
\State Hub node builds the connectivity table for the whole network after collecting all routing tables from other nodes.
\State Hub node \heng{executes Algorithm~\ref{algo:tdma} to build the communication schedule of each node} and floods the schedule out; all the other nodes forward the schedule after receiving it.

\textbf{Data Passing Phase}

\State Each node sends / receives / sleeps based on the schedule.

\textbf{Sleep Phase}
\State All nodes sleep until the next Data Passing Phase.
\EndProcedure
\end{algorithmic}
\end{algorithm}

\subsection{Why centralized mesh protocol?}
\label{sec:why_centralized}

The communication in our mesh network must guarantee there is no collision where two nodes send data to a third node at the same time. Therefore, when creating the schedule for each node to communicate data, we need to understand the whole network structure so as to avoid such communication collisions. Therefore, we design a centralized approach where the hub node in the LoRa mesh is in charge of collecting information from other nodes to understand the whole network structure as well as creates the schedule according to the network structure. It first collects the routing table from each LRMN node so that based on the neighborhood information of each node, it can construct the whole network structure, namely the connectivity table referred in line 5 of Algorithm~\ref{algo:lora_mesh}. Therefore, the scheduling algorithm is centralized and we empirically choose the node closest to the LoRa gateway as the \textit{LoRa hub} to do those jobs. 

\noindent \textbf{Build individual routing table}

\begin{figure}[!ht]
\centering
\includegraphics[width=\linewidth]{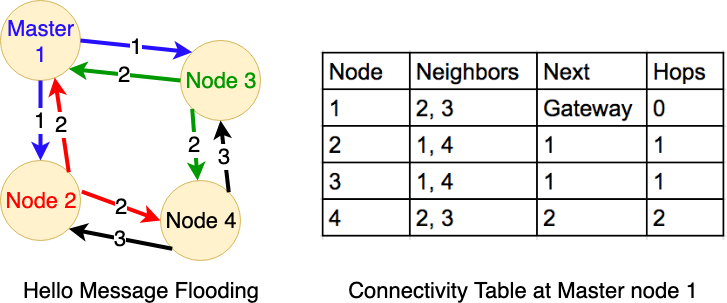}
\caption{Flood hello message to build routing table in each sensor node. After routing tables are built and collected by Node 1, a connectivity table is created to reflect the connectivity of the whole network.}
\label{fig:routing}
\end{figure}

We used a simple version of routing information protocol~\cite{hedrick1988routing} to create the routing table. Figure~\ref{fig:routing} shows how the routing table is built. The LoRa hub node broadcasts a hello message and whoever receives the hello message will rebroadcast it. The hello message includes the information of who the sender is and the shortest distance from the sender to the hub node. Eventually, all nodes will hear the hello message from their neighbors and build an individual routing table. 

\noindent \textbf{Collect individual routing table}

The individual routing tables are sent to the hub to create the connectivity table, which is then referred by the TDMA algorithm to create routing schedules. Every node will broadcast its routing table to its neighbors. Additionally, when a node receives a routing table that it has not received yet will rebroadcast that routing table. Eventually, after all the routing tables are collected at the hub node, the hub node will build a connectivity table that reflects the whole network structure. Figure~\ref{fig:routing} also shows the connectivity table for that mesh network. 

\noindent \textbf{Create schedules for all the nodes}

The hub node will refer to the connectivity table as well as the customized TDMA scheduling algorithm in Section~\ref{sec:tdma} to build the schedule of each node as shown in Table~\ref{tab:schedule}. The schedule is used by each node in the Data Passing phase to either send or receive data. 


\noindent \textbf{Recover from node failure}

In case of a node failure, the associated nodes that normally receive data from the failed node will immediately discover the failure (no downlink communication from the failed node) during the next TDMA cycle, and will switch to reception mode for further instructions. Eventually, all nodes will not transmit anything, and the hub node detects there is a failure. The hub node will issue a reset beacon message and floods it to all node with similar techniques discussed in Section~\ref{sub:time_sync}. After receiving the beacon, the network will repeat the setup phase and recovers from the node failure.



{\footnotesize
\begin{table}[htb]
\footnotesize
\centering
\caption{The schedule built for the network in Figure~\ref{fig:routing}. After timeslot 6, all nodes switch to Sleep mode.}

\begin{tabular}{lllllll}

\hline
\bf Node / Timeslot & \bf 1 & \bf 2 & \bf 3 &\bf 4  & \bf 5 & \bf 6 \\
\hline
1 & Rx & Rx & Rx & Tx & Tx & Tx \\
2 & Rx & Tx & Tx & Sleep & Sleep  & Sleep\\
3 & Tx & Sleep & Sleep & Sleep &  Sleep & Sleep\\
4 & Tx & Sleep & Sleep & Sleep & Sleep & Sleep\\
\hline
\end{tabular}
\label{tab:schedule}
\end{table}	
}

\subsection{Time Synchronization}
\label{sub:time_sync}
\begin{figure}[!ht]
\centering
\includegraphics[width=\linewidth]{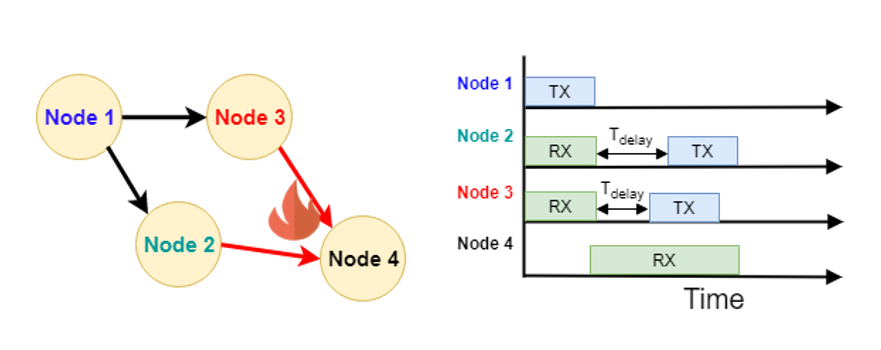}
\caption{Left figure represents the possible package collision caused by time sync with concurrent flooding. Right figure shows the solution of the time synchronization process with random delay.}
\label{fig:time_sync}
\end{figure}

Time accuracy is crucial for any TDMA based collision-free network. Un-synchronized time across the network will result in data loss and network failure. In addition, without an external time source, the micro-controller often relies on crystal oscillator to record time. However, the crystal oscillator drifts over time. Therefore, periodic time synchronization over the network is required for stable operation. Ebi et al~\cite{ebi2019synchronous}. employ an external time source module (GPS) in their network to acquire the coordinated universal time (UTC) at RNs \cite{ebi2019synchronous}. This time will transmit to the connected SNs by a "beacon flooding" with TDMA scheduling. However, using TDMA in a LoRa mesh network for down-link communication will significantly increase the overhead of the network. On the other hand, concurrent flooding addressed the need of the smaller overhead at a cost of higher chances of package collision~\cite{ctlora}. Fig. \ref{fig:time_sync} demonstrates the possible collision that could happen in such approach. To overcome this issue, we insert a random delay between the flooding messages to minimize the possibility of package collision similar to Liao et al.'s work~\cite{ctlora}.

Fig. \ref{fig:time_sync} shows the detailed time synchronization process, for each synchronization cycle, the center node (Node 1 in Fig. \ref{fig:time_sync}) will initiate a 5 bytes beacon package containing source of this beacon ($N_{source}$), number of hops ($N_{hops}$) from the center node and the random delay ($T_{delay}$) before transmitting this beacon from the center node. Any node that receives this beacon will wait for $T_{delay}$ that is smaller than the LoRa symbol time $T_{symbol}$ and then immediately re-transmit this beacon. This approach minimizes the package collision as well as the down-link overhead since the flooding beacon is only 5 bytes. After the beacon has been received by the nodes, the relative time elapsed $T_{past}$ since the previous node's transmission can be calculated as follows:

\begin{equation}
    T_{past} = T_{beacon} + T_{node} +  T_{delay}
\end{equation}

where $T_{node}$ is the time that is need for node to process and re-transmit the beacon, $T_{beacon}$ is the time on air of the package.

However, because of the variability of the $T_{node}$ due to the SPI communication between the micro-controller and LoRa transceiver and the imperfect time synchronization, we manually expand the receiving windows by 5ms to compensate for the inconsistency. This process will synchronize the timing across all the node in the network to avoid the time drifts over a long period of time.

\subsection{Adaptive LoRa Link (ALL)}

One of the benefits of adapting the LoRa technology is the ``degrees of freedom" at the physical layer. Ochoa et al. study suggested that the the potential of an adaptive LoRa solution (i.e., in terms of spreading factor, bandwidth, transmission power, and topology) could greatly optimize energy consumption without sacrificing the communication range\cite{ochoa2017evaluating}. In our proposed LoRa mesh network, we provide the proof of such concept by adopting the Adaptive LoRa Link (ALL) to further improve the energy performance of the network. Although there are many physical parameters that can affect the energy of the LoRa link, SF and transmission power have the most impact in term of the energy consumption. In this work, as a proof of concept, we are only focusing on adjusting the transmission power of the network. However, similar methodologies will apply for adjusting the SF.

During the setup phase, in addition to building a routing table, each node will note the Received Signal Strength Indicator (RSSI) which is a estimated measure of the signal power level from each of the LoRa packages it received. Once the network is stabilized, during the first TDMA cycle, each node will be aware of the RSSI from its previous transmission. Then each node will adjust its $P_{TX}$ based on collected RSSI value. After the adjustment, a new RSSI will be updated to verify the quality of each LoRa link (RSSI $\>$ -120dBm). This process will increase the energy efficiency of each LoRa link without degrading link quality.





\subsection{ANT Hub-and-Spoke Network}

\begin{figure}[!ht]
\centering
\includegraphics[width=0.5\linewidth]{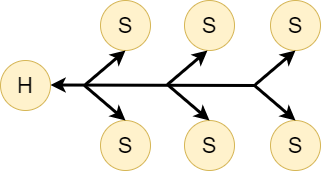}
\caption{ANT shared channel hub-and-spoke network structure. H represents the hub node (Channel Master) and S represents the spoke node (shared slave)}
\label{fig:ant_tree}
\end{figure}

We use ANT to build the Short-Range Network (SRSN). ANT auto shared channel (ASC) is a communication channel specified by ANT~\cite{ant_specification}, and it is used to build reliable bi-directional communication in a hub-and-spoke topology. The ASC communication structure is shown in Figure~\ref{fig:ant_tree}. An ANT hub node receives data from other spoke nodes. \heng{All spoke nodes share a single channel to communicate with the hub. ASC supports up to 66K spoke nodes. By default, ASC requires a user specified channel master node to establish the network, we added a layer on top of that called Adaptive Mode Switching (AMS) to dynamically select the hub node. AMS is based on a round robin (based-on battery level) campaign to select the hub node.
During the setup phase, each node will broadcast its ID and battery level and actively listens for other near-by ANT broadcast,  node with the highest battery level will become the hub node. If multiple nodes have the same battery level, the node with the lowest ID will become the hub node. 
Once the hub node was selected, the ASC uses an ANT proprietary shared channel topology to establish the network, with the hub node being channel master and the rest of the nodes become shared slaves~\cite{ant_specification}. }

\begin{figure}[!ht]
\centering
\includegraphics[width=\linewidth]{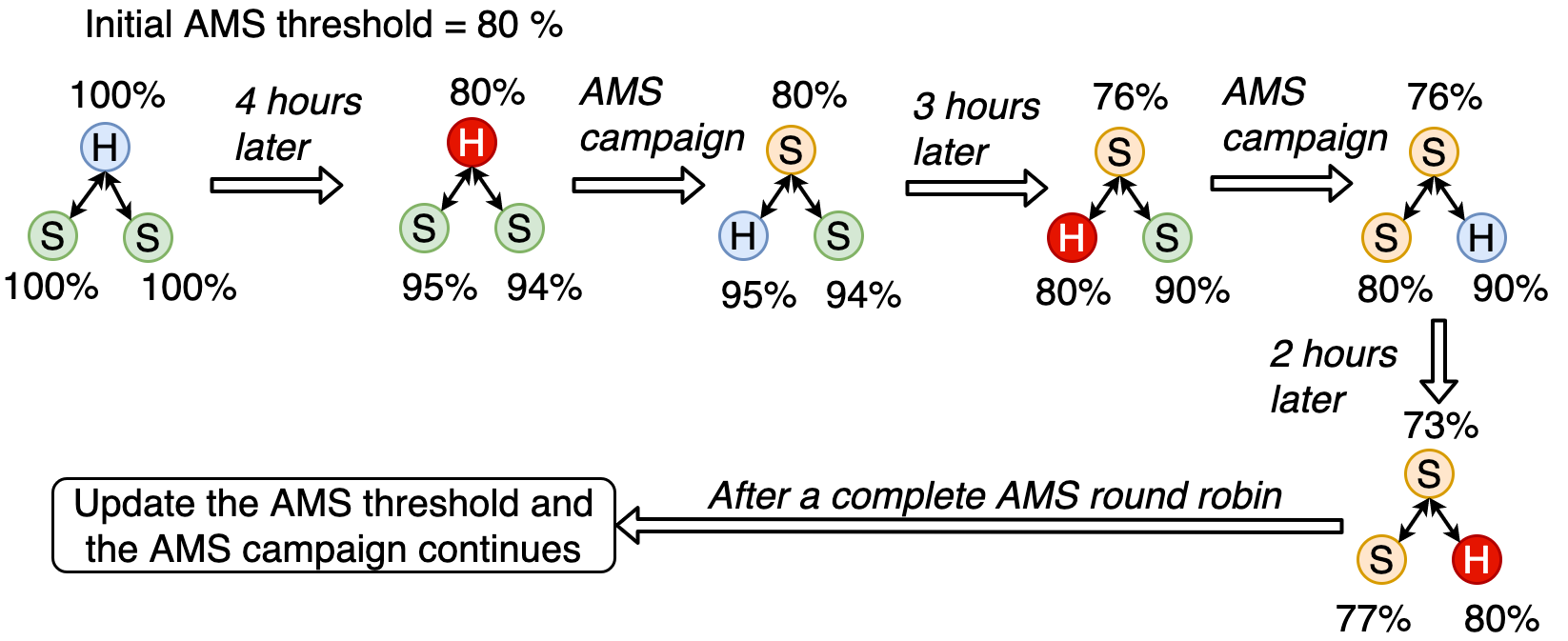}
\caption{Adaptive Mode Switching (AMS) flowchart}
\label{fig:AMS}
\end{figure}

In each SRSN, all the spoke nodes send data to a hub node via ANT and the hub node is in charge of uploading the aggregated data to the cloud. The hub node has two methods to upload the aggregated data. First, if the SRSN is a standalone network and associated with an ANT gateway, it does not need to enable the AMS functionality but directly sends data to the ANT gateway. Second, if the SRSN is part of a LRMN (the SRSN clusters in Figure~\ref{fig:solution_architecture}), the hub node will switch to LoRa mode to route the aggregated data to the LoRa gateway. Clearly, the hub node consumes more energy than the spoke nodes. Therefore, AMS is used for balancing the energy consumption. Figure~\ref{fig:AMS} demonstrates this idea. When a hub node drains to a lower-than-threshold battery level, it will issue a AMS message and the spoke node with highest battery and within the same SRSN will be selected as the new hub. The original hub node then works as a spoke. This AMS feature guarantees no single node in SRSN is significantly drained and ensures the SRSN cluster still remains in the same LRMN network.

\begin{figure}[!ht]
\centering
\includegraphics[width=\linewidth]{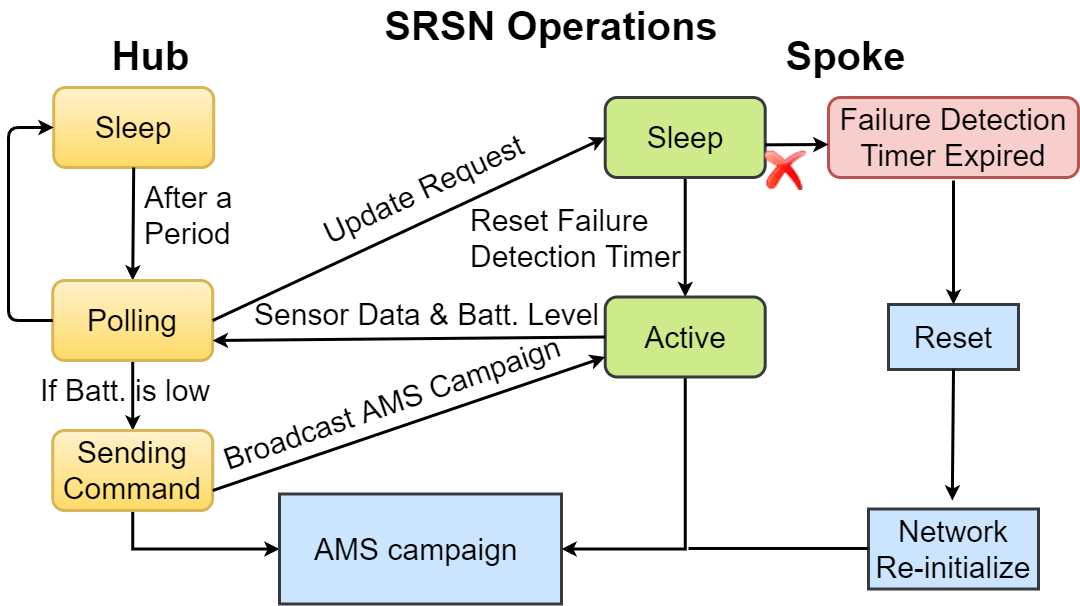}
\caption{\heng{SRSN Normal Operations and Failure Tolerance.}}
\label{fig:master_slave}
\end{figure}

\heng{To address the potential failures of the hub node, we use a heuristic method, a timer (Failure Detection Timer in Fig.~\ref{fig:master_slave}) to detect the failure of the hub node. As shown in the left part of Figure~\ref{fig:master_slave}, during normal operation, the hub node periodically sends an update request to each spoke node to request for new sensor and battery data. When a spoke node finishes data upload, the spoke node will reset the failure detection timer. This timer is set to be much longer than the data upload periodicity (e.g. 5 $\times$ periodicity) to count for any possible data loss. In the scenario of hub node failure (the right part of Fig.~\ref{fig:master_slave}, the timers on the hub nodes will expire and all the spoke nodes will reset itself, resulting in the entire SRSN to be re-initialized. The remaining nodes will form a new SRSN without the failed node. After the new SRSN is initialized, the new hub node will switch on LoRa reception mode and waits for the next TDMA cycle to join back to the LoRa mesh network.}

\subsection{Hardware Sensing Platform in Our Deployment}
There are three main challenges when designing our platform for deployment: building a sufficiently low weight, low cost and energy efficient hardware capable of mass production, incorporating numerous subsystems to facilitate various applications (e.g.smart agriculture and smart city), and protecting the electronics from harsh environmental conditions.

\begin{figure}[!ht]
\centering
\includegraphics[width=\linewidth]{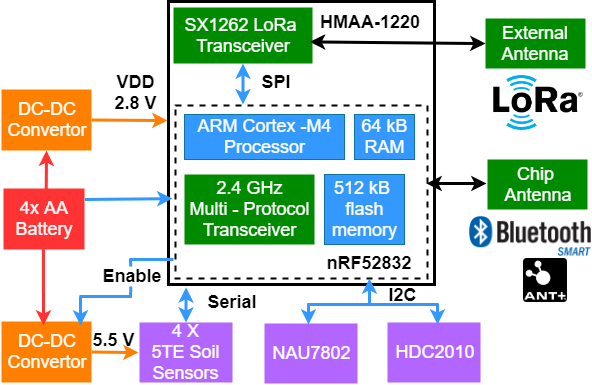}
\caption{\heng{System block diagram of the hardware platform. Adopted from \cite{ieee2019}}}
\label{fig:block}
\end{figure}

\begin{figure}[!ht]
\centering
\includegraphics[width=0.8\linewidth]{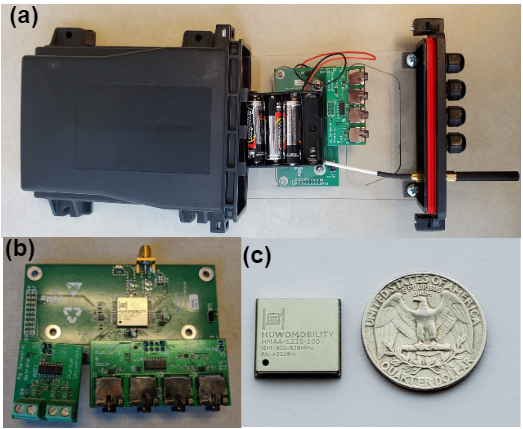}
\caption{The hardware platform: (a) Motherboard PCB and battery holder with its packaging, (b) Motherboard PCB with two daughter boards, (c)the HMAA-1220 module}
\label{fig:hardware}
\end{figure}

The hardware platform used in this study builds on the hardware platform that we reported in our earlier work\cite{ieee2019} as shown in Fig. \ref{fig:block} and Fig.\ref{fig:hardware}(a). It utilizes the HMAA-1220 wireless transceiver module (Fig.\ref{fig:hardware}(b)) from HuWoMobility\cite{huwo}. The HMAA-1220 wireless transceiver was powered by nRF52832 chip from Nordic Semiconductor\cite{nrf} and SX1262 LoRa transceiver from Semtech\cite{semtech}. nRF52832 features a low power 32-bit ARM Cortex-M4F processor with a built-in-radio that operates in the 2.4 GHz ISM band and supports ANT, BLE and Bluetooth 5 wireless protocols with up-to +4 dBm transmit power. It is also equipped with 64 kB RAM and 512 kB of flash storage which can be used for storing data for in-node data analysis.

SX1262 from Semtech \cite{sx1262} is the new sub-GHz radio transceivers which is ideal for long range wireless applications. It supports both LoRa modulation for LPWAN applications and FSK modulation for legacy use cases. In addition, SX1262 also complies with the physical layer requirements of the LoRaWAN specification released by the LoRa Alliance\cite{lorawanspecification} and the continuous frequency coverage of SX1262 from 150 MHz to 960 MHz allows the support of all major sub-GHz ISM bands. SX1262 was designed for long battery life with current consumption of 4.6 mA in active receive mode and 600 nA in sleep mode. With the highly efficient integrated power amplifiers, SX1262 can transmit up-to +22 dBm while having a high sensitivity down to -148 dBm. Along with the co-channel rejection of 19 dB in LoRa mode and 88 dB blocking immunity at 1 MHz offset, SX1262 provides a maximum of 170 dB link budget which is ideal for long distance communication.

Figure\ref{fig:hardware}(a) shows the printed circuit board (PCB) with its packaging. The HMAA-1220 (Fig.\ref{fig:hardware}(b)) module was mounted on a ``motherboard" with 4 LEDs and some pinouts for connecting different ``daughter board" (Fig.\ref{fig:hardware}(c)). This design choice allow us to extend the flexibility of our hardware platform to facilitate different applications. The entire PCB was enclosed in an IP67 Industrial-grade packaging for protection against environmental factors.

%% file: results.tex
\section{Results}
\label{sec:results}

\subsection{In-lab Evaluation}

To verify the functionality and stability of our network, we conducted a series of both in-lab tests as well as real-world deployments.

\subsubsection{Network Stability Test}

We performed laboratory experiments with 9 sensor nodes with 1 node acting purely as standalone receiver monitoring the entire network to check the stability of the network.

\paragraph{Network Setup}

\begin{table}[]
\centering
\caption{LoRa configuration for the in-lab test and field deployment}
\label{tab:test_condition}
\begin{tabular}{llcc}
\hline
\bf Parameter & \bf & \bf In-lab test & \bf Field Deployment\\[2pt]\hline

\begin{tabular}[c]{@{}l@{}}Spreading\\ Factor\end{tabular} & SF & 7 & 7 \\ 
Bandwidth & BW & 125kHz & 125kHz \\ 
\begin{tabular}[c]{@{}l@{}}Preamble\\ length\end{tabular} &  & 8 & 8 \\ 
\begin{tabular}[c]{@{}l@{}}Transmission\\ Power\end{tabular} & $P_{Tx}$ & +0dBm & Variable \\ 
Coding Rate & CR & 4/5 & 4/5 \\ 
CRC checking &  & \begin{tabular}[c]{@{}c@{}}Head \&\\ Payload\end{tabular} & \begin{tabular}[c]{@{}c@{}}Head \&\\ Payload\end{tabular} \\ \hline
\end{tabular}
\end{table}

\begin{figure}[!ht]
\centering
\includegraphics[width=\linewidth]{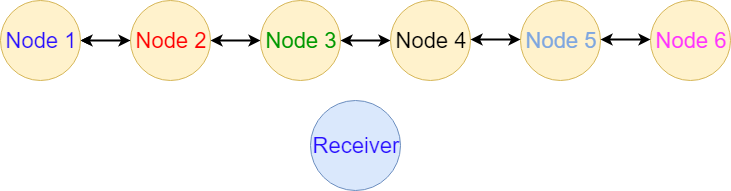}
\caption{Network structure for the in-lab system stability test.}
\label{fig:labtest}
\end{figure}

Fig. \ref{fig:labtest} shows the configuration of the network structure. All 8 nodes are placed together on a lab bench along with the receiver. Node 1 is the center node for this test. Testing the network structure in this condition is nontrivial since all sensors are in close proximity, the long-range capability of LoRa will not able to form the structure that we desired since all node are in range with each others no matter how we configure the LoRa parameters. To resolve this issue, we created a filter in the low-level firmware (LoRa driver) to block connections from un-wanted sensor nodes. For example, in the structure shown in Fig \ref{fig:labtest}, Node 4 should only receive data from Node 3 and Node 5 in the desired structure. However, in reality node 4 is able to receive data from all nodes because of the close proximity between sensors. The firmware filter will filter any data from nodes other than node 3 and node 5. All other data will be ignored to simulate the desired network structure. The biggest benefit of the approach is that the mesh-layer of the network is completely un-touched, meaning from a network stack point of view, this lab test will be able to simulate the real-world conditions. 

Table \ref{tab:test_condition} shows the LoRa configuration of the lab experiment. To fully review the performance as well as the stability of the network, we kept the network running with each node programmed to transmit one 64 Bytes data package per minute (1.8kBps) for an entire week. As the performance indicator, the receiver records all of the network traffic from all nodes. RSSI are not evaluated in this test since we intentionally lowered the TX power of all LoRa nodes. We evaluated the reliability and stability of data packet delivery for each individual node using the packet deliver rate (PDR), i.e. the ratio between the number at the center node (\# RECEIVED) and the number of packets that should have been received (\# EXPECTED). With the help of the traffic monitoring receiver, two sets of PDRs were can be calculated, $PDR_{i}$ and $Total\_ PDR_{d}$:


\begin{equation}
PDR_{i} = \frac{\sum\#RECEIVED_{i}}{\sum\#EXPECTED_{i}}
\end{equation}

Where $PDR_{i}$ is the packet delivery rate of node \textit{i} during the entire week.

\begin{equation}
Total\_ PDR_{d} = \frac{\sum_{d}\#RECEIVED}{\sum_{d}\#EXPECTED}
\end{equation}

Where $Total\_ PDR_{d}$ is total the packet deliver rate of all the node during the \#dth day.

The estimation of the number of expected packets are based on each specific node with constant transmission interval (1 min). Because of the nature of multi-hop network, counting the number of packets arriving at the receiver will counts for both node-specific performance as well as the multi-hop route from that specific node to the center node. In contrast, $Total\_ PDR_{d}$  provides an overview of the network stability over time. Instead evaluating node-specific parameters, $Total\_ PDR_{d}$ provides an overview of the system stability of time synchronization as well as the TDMA routing.

\begin{figure}[!ht]
\centering
\includegraphics[width=0.9\linewidth]{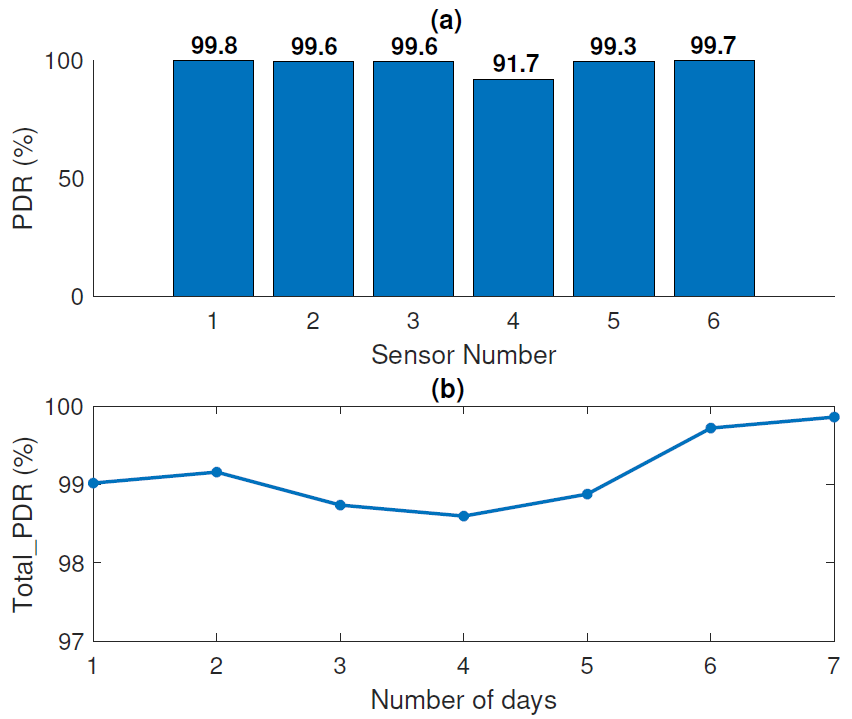}
\caption{Results (PDR and $Total\_PDR$) for the in-lab system stability test.}
\label{fig:labtest_result}
\end{figure}

Fig. \ref{fig:labtest_result}(a) shows the results of the $PDR_{i}$ of our one-week test, all nodes show more than 99$\%$ PDR except for node 4. Fig. \ref{fig:labtest_result}(b) shows the results of the $Total\_ PDR_{d}$ for the same test, it suggests our network stability is very strong and is not time dependent.

\subsubsection{Power Consumption}



\begin{figure}[!ht]
\centering
\includegraphics[width=\linewidth]{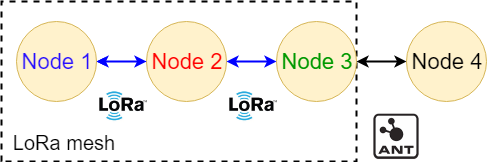}
\caption{Network setup for the energy consumption test}
\label{fig:power_test}
\end{figure}

The power consumption was measured for one node under real-life conditions for a network structured with four nodes as shown in Fig \ref{fig:power_test}. In addition, each sensor is connected with HDC2010 temperature and humidity sensor. For each cycle, each node will transmit a package of 64 bytes including one temperature and humidity reading. Due to the nature of our mesh network, energy consumption within the network varies depending on i) the position of the participating node in the hierarchy of the mesh network, and ii) the topology type of the network. As shown in Fig. \ref{fig:power_test} node 1, 2, and 3 are connected with the proposed LoRa mesh protocol. Node 3 also communicates with node 4 via ANT. The goal of this configuration is to route the data from Node 4 (ANT), Node 3 (ANT + LoRa), and Node 2 (LoRa) to Node 1 (LoRa). For each TDMA scheduling cycle, Node 4 will transmit one package via ANT to Node 3 and Node 3 will forward this package plus its own package to Node 2 via the LoRa. Then, Node 2 will transmit 3 packages (2 received + 1 own) to Node 1. The LoRa configuration for this test is identical to the field deployments with +18 dBm transmission power.  

\begin{figure}[!ht]
\centering
\includegraphics[width=\linewidth]{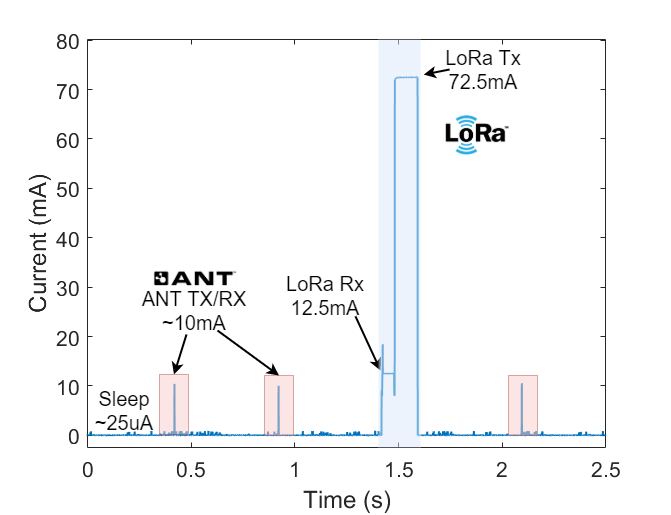}
\caption{Measured current profile for Node 3 during one transmission cycle where blue region represents LoRa's Tx and Rx windows and red region represents the ANT transmission window.}
\label{fig:labtest_battery}
\end{figure}

Power consumption was measured using a N6705B DC power analyzer from Agilent Technology\cite{N6705B}. Each sensor node is powered with 3.3V DC by the DC power analyzer. Fig. \ref{fig:labtest_battery} shows part of the current profile of Node 3 where each state of operation is clearly marked. The hardware consumes around 25 $\mu $A during sleep, 10 mA during ANT TX, 12.5 mA during LoRa RX and 72.5 mA during LoRa TX. Therefore, the average current consumption with 10 minutes TDMA cycle delay for Node 3 is 56 $\mu$A which can be translate to 5.2 years of expected battery life with standard AA alkaline batteries (2500 mAh). Table \ref{tab:batterylife} shows the average current consumption and the expected battery life (with 2 AA alkaline batteries) of each node. Comparing between nodes 1,2, and 3, the energy consumption of different nodes is determined by the number of receive/transmits windows. Note that Node 4 consumes significantly lower power. This shows that ANT radio is superior in terms of energy efficiency compared with LoRa. In conclusion, even though energy consumption is highly dependent on the network structure and while some nodes do consume high power, our network still has a very acceptable expected battery life across all nodes. 


\begin{table}[]
\centering
\caption{Measured Power profile and Expected Battery Life with 2xAA Battery}
\label{tab:batterylife}
\begin{tabular}{lcc}
\hline

 & \multicolumn{1}{l}{Average Current Draw ($I_{avg}$)} & \multicolumn{1}{l}{Expected Battery Life} \\\hline
Node 1 & 33 $\mu A$ & 9 years \\
Node 2 & 74 $\mu A$ & 4 years \\
Node 3 & 56 $\mu A$ & 5 years \\
Node 4 & 25 $\mu A$ & 11 years \\\hline
\end{tabular}
\end{table}

\begin{figure}[!ht]
\centering
\includegraphics[width=\linewidth]{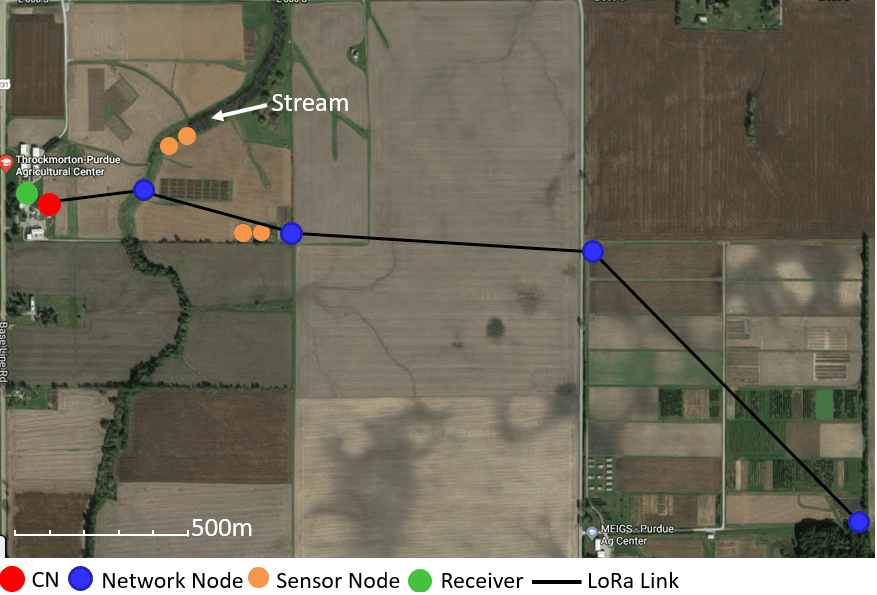}
\caption{Deployment of the proposed mesh network on Purdue University West Lafayette campus. Each blue dot represents a LoRa node and the black dash line represent a stable LoRa link}
\label{tpac_mesh}
\end{figure}

\begin{figure*}[h]
\centering
\includegraphics[width=0.8\linewidth]{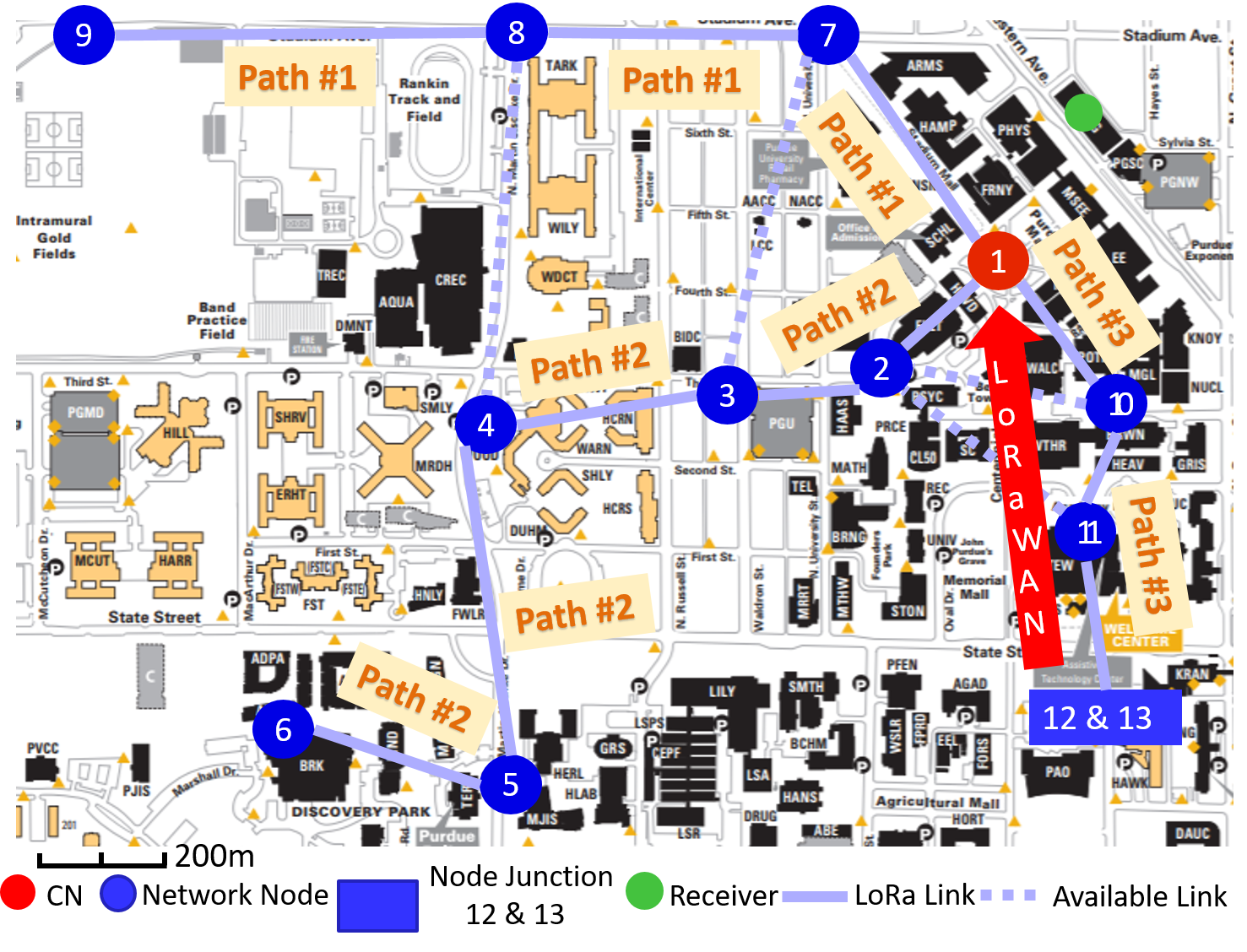}
\caption{Map of the network deployment on Purdue campus with 13 nodes. The red dot represents the center node  (CN), the network nodes are represented as blue dots, the blue square represents the node junction (Node \#12 and \#13) and the green dots represents the receiver.}
\label{fig:purdue_deployment}
\end{figure*}

\begin{figure*}[htp]
\centering

\subfloat[]{%
  \includegraphics[width=0.9\linewidth]{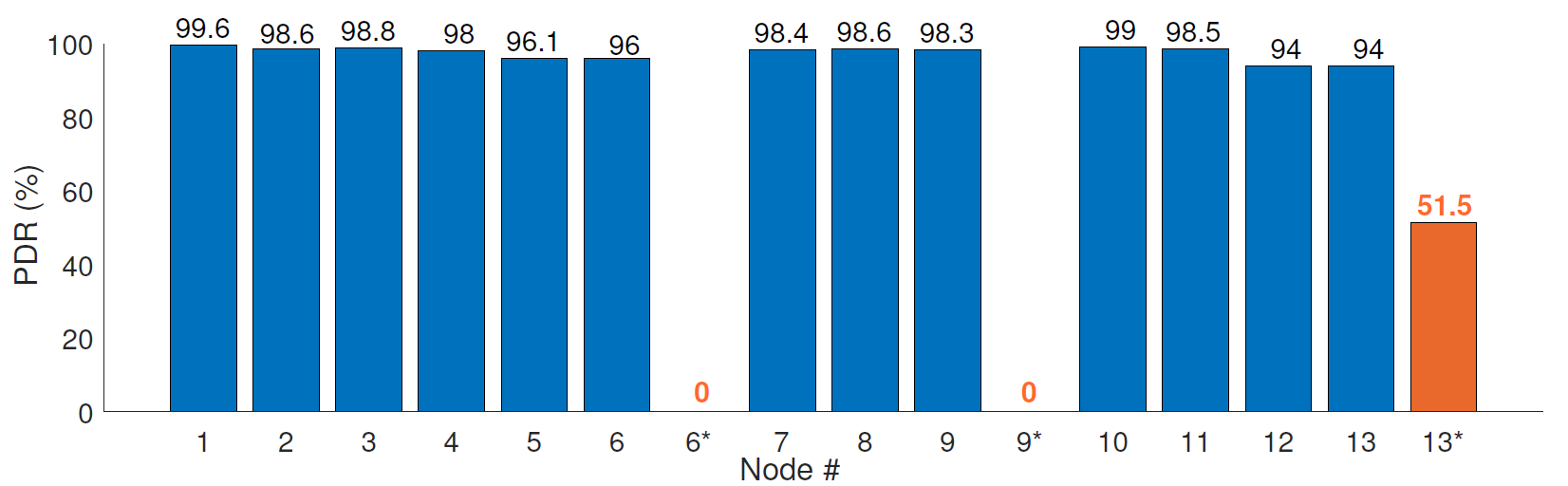}%
}

\subfloat[]{%
  \includegraphics[width=0.9\linewidth]{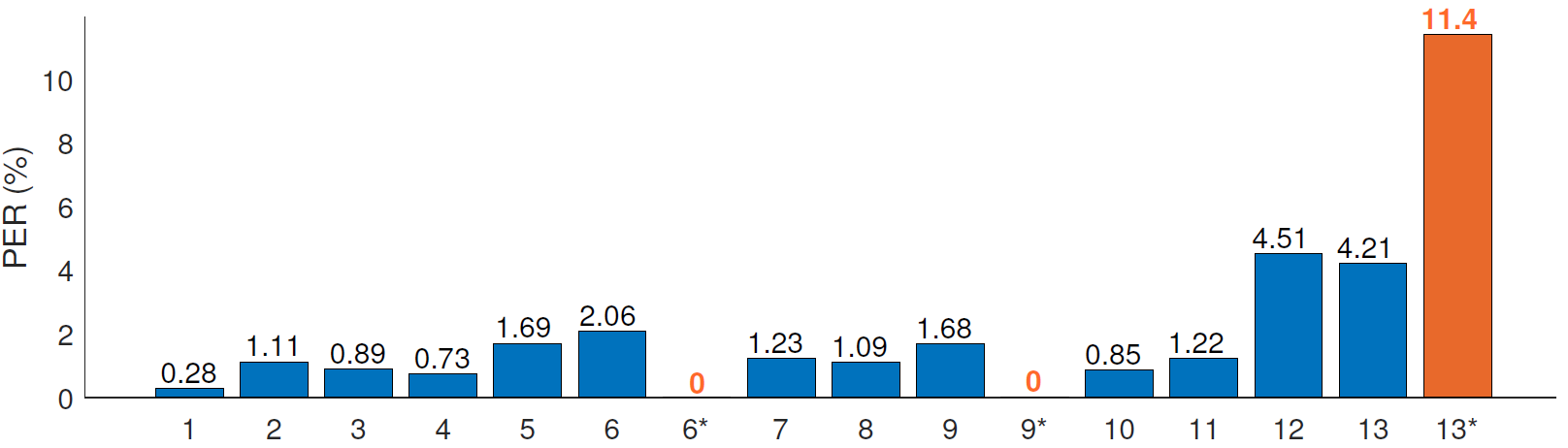}%
}

\subfloat[]{%
  \includegraphics[width=0.9\linewidth]{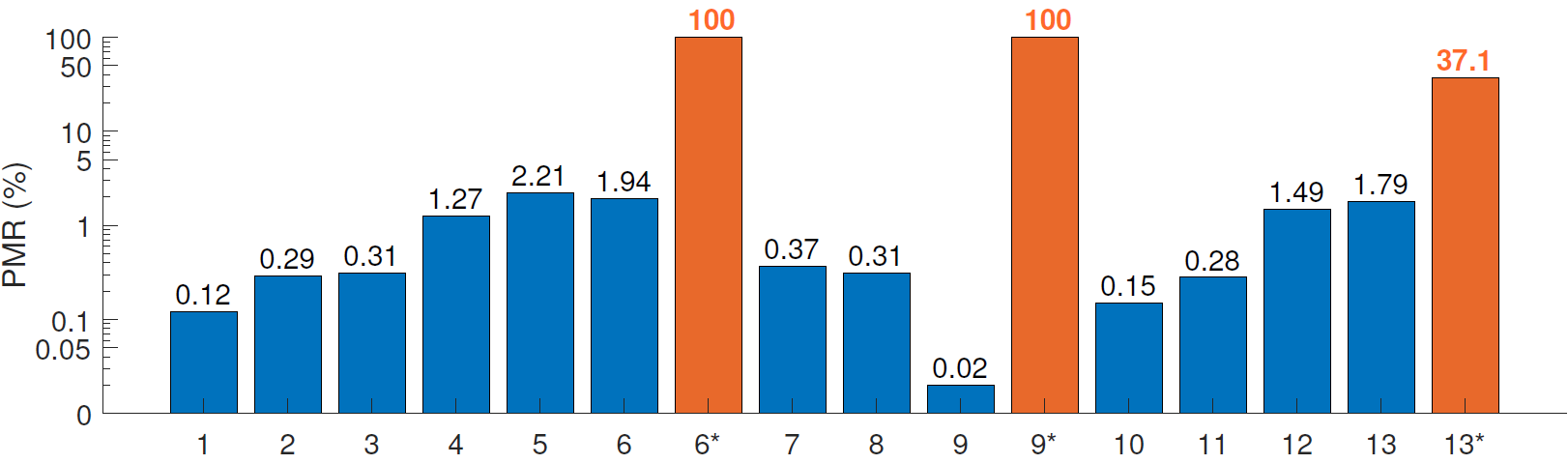}%
}

\caption{(a) Packet Delivery Rate (PDR), (b) Package Error Rate (PER), (c) Package Miss Rate (PMR) of the two weeks on-campus deployment. Blue bars represent node 1 to 13 across the network. Orange bar represents the SF12 LoRaWAN reference from node 6, 9 and 13 respectively}
\label{fig:campus_results}
\end{figure*}

\subsubsection{Hardware durability}


To fully test the performance and durability of our hardware platform against harsh environments, we deployed 4 units equipped with our smart agriculture interface at Throckmorton-Purdue Agricultural Center (TPAC) at Purdue University. All four units were equipped with temperature and humidity sensors to monitor the environmental conditions at the farm and were placed 1 meter above the ground. The deployment locations are shown in Fig. \ref{tpac_mesh}. Two of the units with printed thin-film nitrate sensors were installed in the stream to measure the nitrate pollutants from fertilizer runoffs in the stream. The other two units were interfaced with four independent ECH2O 5TE Soil sensors to monitor the soil temperature, conductivity, moisture, and dielectric at four different depths\cite{meter}. All four units were programmed to send data every 10 minutes and the data received at the receiver is uploaded to the data server and displayed on the web. As of January 2020, these units have been continuously operating for more than a year without major failure. The highest temperature recorded is above 110 $^{o}F$ and the lowest temperature is -40 $^{o}F$. This proves our hardware is capable of withstanding harsh environmental conditions while being in situ in an unmonitored outdoor environment.


\subsection{Large-scale Deployment}
 

\begin{figure*}[h]
\centering
\includegraphics[width=\linewidth]{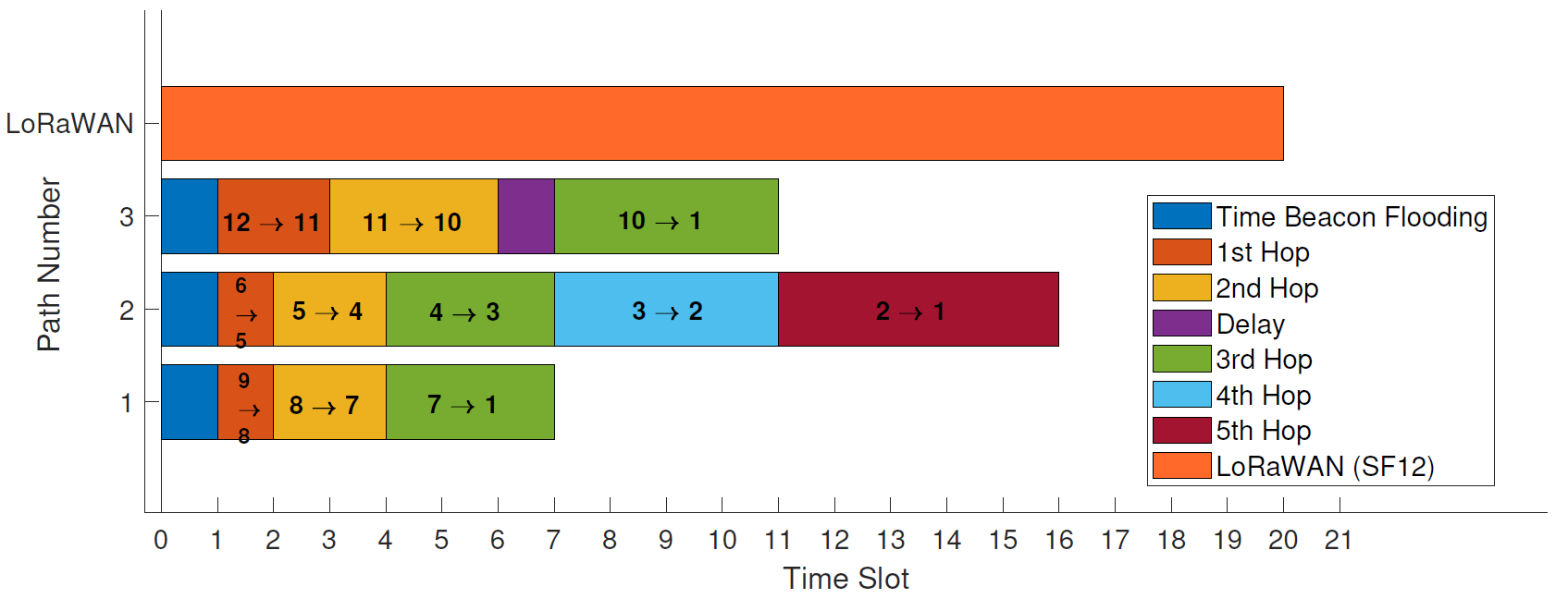}
\caption{TDMA schedule of the campus deployed network, each Hop is clearly marked. Path 1 to 3 corresponds to the path from Fig. \ref{fig:purdue_deployment}. The total length represents the total overhead of the corresponding path.}
\label{fig:overhead}
\end{figure*}

\subsubsection{Farm deployment}

Our proposed network was first tested in TPAC to evaluate the multi-hop performance of the mesh network for covering long-ranges. Five mesh nodes were deployed in linear hop formation in addition to the existing 4 sensing nodes equipped with soil moisture sensors and flexible nitrate sensors. The newly deployed mesh nodes were equipped with temperature and humidity sensors as well as nitrate sensors and powered with 4 AA batteries. Fig. \ref{tpac_mesh} shows the map of the farm deployment: The green dot represents the receiving computer; The red dot represents the center node of the mesh network; The blue dots represents the 5 mesh nodes; The yellow dots represent the four previously deployed sensor nodes that were not part of the mesh network. The LoRa configuration of each node is shown in Table \ref{tab:test_condition} with SF7 and TX power = 15 dBm. Each node was programmed to send one 64-byte packet every 10 minutes. Once each cycle is complete, the center node will forward the packages to the receiver for upload. With 4 linear hops, the proposed mesh network is able to cover 3 km in farmland with more than 98\% PDR across all nodes with SF7. This experiment confirmed that our mesh network is able to cover long distances with low spreading factor (SF7).

\begin{figure}[h]
\centering
\includegraphics[width=\linewidth]{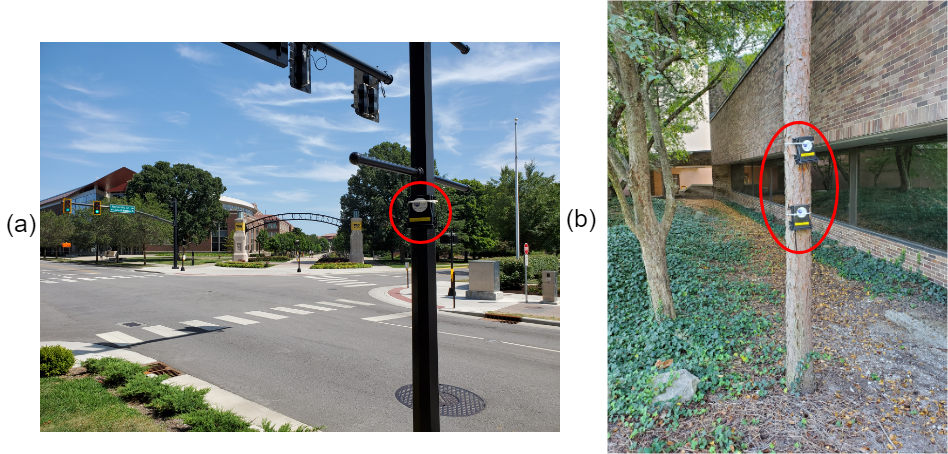}
\caption{Deployment at Purdue campus. (a) node \#7 installed on a street lamp post (b) node junction consisting node 12 and 13 next to a campus building}
\label{fig:purdue_deploy}
\end{figure}

\subsubsection{Campus deployment}

In the campus-scale deployment, we placed 13 LoRa nodes, distributed in a 1.1 km by 1.8 km area of Purdue University campus. All 13 nodes were deployed and continuously operated for a period of two weeks.  Fig. \ref{fig:purdue_deployment} shows the the complete map of our campus-scale experiment where a complete mesh network is established. Each blue dot represents a network node which are randomly and evenly distributed across the entire Purdue campus (Fig. \ref{fig:purdue_deploy}(a)). The red node represents the location of the center node. The green dot represents the receiver, a laptop connected with $SX1272DVK1CAS$ (LoRa development kit) from Semtech\cite{semtechdk}. Each solid line represents the actual LoRa link for the deployed network and the dotted line represents available LoRa links that were not being used. The node junction represents two nodes (node 12 and node 13) that were deployed on purpose at close range and were communicating via ANT instead of LoRa (Fig.\ref{fig:purdue_deploy}(b)). In addition, three paths that are highlighted in Fig. \ref{fig:purdue_deployment} represent three distinguished data flows that are formed by the mesh network. For instance, path number 1 represents the path 9 $\rightarrow$ 8 $\rightarrow$ 7 $\rightarrow$ 1. All nodes (including the center node) are located 1 meter above ground level as shown in Fig. \ref{fig:purdue_deploy} and the receiver is placed on the 3rd floor inside of an office building facing south-east. The furthest node is placed 1.5 km away from the receiver and across more than 18 buildings in between.

Each node was programmed to transmit 64 Byte package at a fixed time interval (2 minute) with the LoRa configuration shown in table \ref{tab:test_condition}. In addition, two furthest nodes 6 and 9 will send an additional package with SF12 outside of the TDMA cycle as comparison with transitional ALOHA based network such as LoRaWAN. Furthermore, since the LoRa radio of node 13 in the node junction was not active (it only communicates to node 12 via ANT), it was enabled to broadcast (in parallel with the mesh network, acting like a LoRaWAN node) with SF12 directly  to the receiver used as the comparison for evaluation.  After each transmission cycle, the center node will forward all incoming packages to the receiver which will upload all data to the cloud for future analysis. All data packages are checked with 16-bit Cyclic Redundancy Check(CRC) to verify their integrity, corrupted data package will be marked and stored. Fig. \ref{fig:overhead} shows the TDMA schedule of the deployed network with each slot set to 125 ms for transmitting a 64 Byte package. The three paths represent the three continuous LoRa links as shown in Fig. \ref{fig:purdue_deployment}. Each section represents one different LoRa hop and are color coded for clarity. The delay inserted after the 3rd hop is required to avoid package collision with node 7 from path 1. SF12 reference represents the time it took for one 64 Bytes transmission with SF12 as comparison. It is clear that even for the longest path (Path \#2) with 5 hops in between, the overhead of the furthest mesh node (Node 6) is smaller than the single-hop LoRaWAN node. However, it is worth noticing that the overheads are greater for the nodes that were closer to the center node (Node 7, 10, and 2) and it is a necessary trade-off between power efficiency and network robustness.

Over the entire evaluation period of two weeks, each node transmitted one 64 Byte package every two minutes, which means a total of 9360 packages were expected from each node (13 days of operation were considered, network was taken down by one day for evaluation). As in the previous section, the network integrity is evaluated by analyzing PDR of all nodes. In addition, the Package Miss Rate (PMR) and Package Error Rate (PER) are analyzed as well. Similar to PDR, PER is calculated based on the number of the package that were marked as corrupted and PMR are calculated based on number of missed packages:

\begin{equation}
PER_{i} = \frac{\sum\#ERROR_{i}}{\sum\#EXPECTED_{i}}
\end{equation}

\begin{equation}
PMR_{i} = \frac{\sum\#EXPECTED_{i} - \sum\#RECEIVED_{i}}{\sum\#EXPECTED_{i}}
\end{equation}

Fig. \ref{fig:campus_results} shows the end-to-end PDR, PMR and PER based on the total expected number of packages and received packages for all 13 nodes including the SF12 reference, where the blue bars represent all nodes in the mesh network, and the orange bar represents the SF12 reference. Over the entire deployment period of two weeks, the proposed mesh network achieves more than 96\% PDR except for node 12 and node 13 comparing to 51.5\% of the SF12 reference node. In addition, from Fig. \ref{fig:campus_results} (b) and (c), both the PMR and PER for all nodes are significantly lower than the SF12 reference node. This confirms that the proposed network provides much better quality of service particularly for large area networks. Where higher spreading factors are necessary for transitional star network to cover such as in the ALOHA protocol in LoRaWAN's approach, it is worth noticing that the PMR increases as the number of hops increases, this is expected since the time flooding will degrade as the number of hops increases. Imperfect time synchronization will cause time slot mismatch and therefore results in either missed packages (TX/RX window miss match) or package collision (TXs windows miss match). However, from the observation based on the results from Fig. \ref{fig:campus_results}, although both of the PMR and PER did trend to increase as number of hops increased, these effects are minimal comparing with the PDR. Furthermore, node 12 and 13 showed much higher PER ($>$ 4\%), we suspect that this might be due to higher interference in the sub-GHz ISM band since both of those two nodes were located in one of the most populated areas on campus. On the other hand, the LoRaWAN reference node 13* shows a very low PDR (51.5\%) and very high PER and PMR, when compared to our mesh node deployed at the same location (node 12). This confirms that our mesh is able to provide better quality of service which further supports our purposed network structure.  


%% file: limitation.tex
\section{Limitations}
\label{sec:limitation}

As Ochoa et al.~\cite{ochoa2017evaluating} point out, the energy consumption of LoRa mesh nodes can be further optimized by exploiting different radio configurations and the \textit{network topology} (e.g., the number of hops, the network density, the cell coverage). For sparse networks, higher SF is necessary along with higher transmitted power. The Adaptive LoRa Link in our implementation did not include the functionality to change the spreading factor as the network topology in our deployment did not change over time. One aspect of our future work is to include such adaptivity in our implementation.

Another important limitation of the network occurs during the flooding for the TDMA scheduling. In our current configuration, one TDMA schedule table for the entire network is flooded to each node for the simplicity of the design. However, there are two inherent problems with this approach. First, flooding the entire table requires transmitting multiple LoRa packages throughout the entire network. Not only is this approach inefficient, but it also increases the overhead for the setup phase. Second, due to the limitation of the maximum data package (255 bytes at SF7) of LoRa, the maximum number of nodes in the network will be limited. Although, this limitation can be patched with flooding multiple schedule tables throughout the network, this approach will still be inefficient and will significantly impact the overhead during the setup phase. For our future work, instead of flooding a entire network table to every node, we will divide the table, based on the routing path. Only the necessary TDMA schedule will be flooded to each multi-hop path in the network. This approach will significantly reduce the overhead time during the setup phase without sacrificing the network performance.

While LoRaWAN allows for AES 128-bit encryption,
in the current phase of our work, no encryption mechanisms have been deployed. While we plan on deploying AES-128 encryption in future LoRaWAN deployments, a caveat of introducing secure network channels will be the reduction of the available payload size, which may further limit the number of supportable nodes in a sub-network. Thus, we will look into deploying lightweight encryption mechanisms for IoT devices, such as ACES~\cite{benchIoT-2019, ACES2018}. 

As of the sensor data management, there are several other challenges about network latency and data analytics, which we have more thorough discussion in~\cite{technical_report}

%% file: conclusion.tex
\section{Conclusion}
\label{sec:conclusion}

The recent advancement of the Internet of Things (IoT) enables the possibility of data collection from diverse environments using IoT devices. However, despite the rapid advancement of low-power communication technologies, the deployment of IoT network still faces many challenges. In particular, large-scale WSN such as digital agriculture and smart and connected cities remains a major challenge in terms of communication range, quality of service and power consumption.  

This paper presents the design of a hybrid LPWAN mesh network for IoT application that delivers several-kilometers with only low-power nodes while provides excellent QoS. Our work addresses the development of large-scale WSN that is suitable for distinct application areas with real world deployments. To enable the data collection with varying sensors as well as to support wide area coverage with low energy consumption, we designed and manufactured our own sensor node integrating a micro-controller, wireless communication interfaces, and a hybrid network with short (2.4GHz) and long-range (915MHz) communication links.

With our hybrid mesh network, we have shown a significant improvement in both power consumption as well as communication range while comparing with traditional single hop network like LoRaWAN. In addition, full-scale real-world experiments on both Purdue Campus and agricultural farms with more than 20 nodes further suggested that the proposed network significantly improves the quality of service while maintain long-term stability. We provide several areas of future work motivated by our design and experiments on these large scale IoT testbeds, including sophisticated anomaly detection, on-device computation, and network synchronization. 

%% file: acknowledgement.tex
\section*{Acknowledgment}
\label{sec:ack}

The work described in this paper is part of a project funded  through the Wabash Heartland Innovation Network (WHIN) and the SMART Film consortium at Purdue University.

\ifCLASSOPTIONcaptionsoff
  \newpage
\fi